\documentclass[%
 reprint,
superscriptaddress,
 amsmath,amssymb,
prb,
]{revtex4-2}

\usepackage{graphicx}
\usepackage{dcolumn}
\usepackage{bm}
\usepackage{xcolor}
\usepackage[normalem]{ulem}

\newcommand{\pto}{ PbTi$\text{O}_3$ }

\newcommand{\pmn}{ Pb$(\text{Mg}_{1/3}\text{Nb}_{2/3})\text{O}_3$ }

\newcommand{\bnt}{
$(\text{Na}_{0.5}\text{Bi}_{0.5})\text{TiO}_3$ }

\begin{document}

\preprint{APS/123-QED}

\title{Polar vibrational excitations and glass-like response in relaxor PMN}

\author{Kehan Cai}
\affiliation{%
 Department of Chemistry, Princeton University, New Jersey 08544, U.S.A
}%

\author{Pinchen Xie}
\affiliation{
 Program in Applied and Computational Mathematics, Princeton University, New Jersey 08544, U.S.A
}
\affiliation{Applied Mathematics and Computational Research Division, Lawrence Berkeley National Laboratory, Berkeley, California 94720, USA}

\author{Yifan Li}
\affiliation{%
 Department of Chemistry, Princeton University, New Jersey 08544, U.S.A
}%

\author{Roberto Car}
\thanks{Contact Author: rcar@princeton.edu}
\affiliation{%
 Department of Chemistry, Princeton University, New Jersey 08544, U.S.A
}%
\affiliation{
 Program in Applied and Computational Mathematics, Princeton University, New Jersey 08544, U.S.A
}%
\affiliation{%
 Department of Physics, Princeton University, New Jersey 08544, U.S.A
}%
\affiliation{
 Princeton Materials Institute, Princeton University, New Jersey 08544, U.S.A
}%

\begin{abstract}
Relaxor ferroelectrics retain crystalline order while exhibiting glass-like dielectric and thermal responses. Using atomistic models trained on first-principles data, we investigate these responses in the prototypical relaxor \pmn (PMN). Compositional disorder couples elastic motion to electric polarization, producing a continuum of polar vibrational excitations, unlike the discrete polar optical modes of conventional ferroelectric \pto. The simulations qualitatively reproduce the temperature dependence of the low-frequency susceptibility, while the calculated far-infrared spectra agree closely with experiments. At low frequencies, disorder in PMN generates a boson peak comprising excess phonon-like modes and quasi-localized excitations. The latter exhibit a quartic frequency dependence of their density of states and pronounced anharmonicity, as in structural glasses. These excess modes account for PMN's anomalous behavior in the specific heat for temperature higher than 5 K. These results connect the dielectric, optical, and thermal anomalies of a crystalline relaxor through its disorder-induced polar excitation spectrum.
\end{abstract}
\date{\today}

    \maketitle


\section{Introduction}

Relaxor ferroelectrics are compositionally disordered crystalline materials whose dielectric response differs qualitatively from that of conventional ferroelectrics. In prototypical relaxors such as \pmn(PMN) and \bnt(BNT), the compositional disorder arises from the coexistence of different atomic species on the A or B sublattice in the ABO$_3$-type perovskite structure. In these materials, the dielectric susceptibility exhibits a broad, frequency-dependent maximum and a progressive loss of ergodicity upon cooling~\cite{smolenskii1961ferroelectrics,burns1983glassy,viehland1990freezing,levstik1998glassy,cowley2011relaxing}. At low temperatures, PMN also exhibits anomalous thermal properties, including a prominent peak in the reduced specific heat~\cite{tachibana2009thermal}. These observations raise the question of how compositional disorder and its interplay with spatially nonuniform local polarization determine the unique dielectric, optical, and thermal responses in relaxors.

Many existing microscopic descriptions of relaxor behavior have focused on polar nanoregions (PNR), which are nanoscale regions of correlated, almost frozen local polarization that emerge upon cooling~\cite{wakimoto2002ferroelectric,blinc2003field,xu2004neutron,jeong2005direct,gehring2009reassessment,fu2009relaxor}. 
Qualitatively, PNR resemble frustrated spins in magnetic spin glasses. 
Indeed, reduced dipolar-glass models and atomistic simulations can reproduce the progressive freezing of local polar structures~\cite{akbarzadeh2012finite,shin2005development, grinberg2009molecular,takenaka2017slush} and provide a framework for interpreting the dielectric relaxation in relaxors~\cite{viehland1991glassy}. Nevertheless, a phenomenological description based on quasi-static local polar structures does not, by itself, elucidate the microscopic mechanisms by which compositional disorder~\cite{chen1989ordering, randall1990classification, viehland1991glassy, davies2000chemical, akbas2000thermally, xu2000direct, farber2003influence, fu2009relaxor, cabral2018gradient} modifies the organization of collective vibrational excitations that carry polarization and energy.

Understanding the connection between microscopic excitations and glass-like macroscopic responses in relaxors is particularly important because relaxors are at the intersection of conventional ferroelectrics and structural glasses.  
In a compositionally ordered ferroelectric such as \pto (PTO), acoustic phonons are nonpolar. The coupling to a uniform electric field is concentrated in a small number of zone-center polar optical modes. In contrast, structural glass typically hosts excess low-frequency excitations, including quasi-localized excitations (QLE). QLE are non-phononic modes with a core of large atomic displacements coupled to weaker surrounding displacements. The vibrational density of states (VDOS) of QLE follows the characteristic $\mathcal{D}(\nu)\propto\nu^4$ scaling~\cite{laird1991localized,kapteijns2018universal,lerner2021low}. Such excess excitations are related to universal low-temperature anomalies in glasses~\cite{zeller1971thermal, phillips1981amorphous, martin1979specific, meschede1980specific, kofu2024magnetic}. Relaxor materials cannot be assigned unequivocally to either category: they remain crystalline yet possess a quenched compositional structure with extensive short-range chemical order that simultaneously couple elastic, electric-dipolar, and compositional degrees of freedom. How these coupled degrees of freedom reshape the vibrational excitations and how they collectively produce the dielectric and thermal anomalies are the central questions addressed in this work.

Here, we investigate prototypical relaxor PMN using accurate atomistic models based on machine-learned interatomic interactions and polarization trained on first-principles calculations. The compositional disorder is modeled by randomly mixing atomic blocks of fixed nanometer size, representing regions with and without short-range chemical order~\cite{fu2009relaxor, cabral2018gradient}.
We first show that the model reproduces essential signatures of relaxor behavior, including PNR, ergodicity breaking, a diffuse dielectric transition, and the temperature dependence of the dynamic susceptibility. We then demonstrate that compositional disorder induces a coupling between elastic and electric-dipolar degrees of freedom and simultaneously polarizes all vibrational modes. Thus, the system responds to electric perturbations across a broad range of frequencies in the far-infrared (far-IR) region ($10 - 400$ $\mathrm{cm^{-1}}$) and the microwave region ($0.01-10$ $\mathrm{cm^{-1}}$), in stark contrast to ferroelectric perovskites such as PTO, which contain only a finite number of zone-center optical modes that are IR active. These results elucidate the distinctive dielectric response of relaxors~\cite{PhysRevB.96.174113, hlinka2006infrared} from an atomistically resolved, dynamic perspective.

Additionally, we identify an excess of low-frequency excitations in PMN, including QLE exhibiting pronounced anharmonicity and a VDOS that scales quartically with frequency.  These modes lead to a broad and prominent peak in the reduced VDOS ($\mathcal{D}(\nu)/\nu^2$) of PMN, i.e., a boson peak~\cite{mizuno2017continuum}.
We further predict vibrational specific heat in PMN and PTO for $T>5$ K in quantitative agreement with experiments~\cite{tachibana2009thermal}. We can identify the thermodynamic footprint of the boson peak~\cite{phillips1981amorphous, buchenau1984neutron, yu1988low, wuttke1995fast, wischnewski1998sound, talon2002low, ramos2002boson, chumakov2004collective} as a pronounced peak in the reduced specific heat ($C\left(T\right)/T^3$) of PMN. We also identify a weak peak in the reduced specific heat of PTO, which results solely from the non-linear dispersion of acoustic modes near the zone boundary. 

Together, these results connect the dielectric, optical, and thermal anomalies of PMN to the unusual polar excitation spectrum generated by compositional disorder. 
 
The remainder of this paper is organized as follows. Sec.~\ref{sec:machine_learning_models} introduces the machine-learning models of interatomic interactions and polarization in PMN. Sec.~\ref{sec:structural_model_and_PNRs} describes the structural model of compositional disorder and the emergent local polar structures. Sec.~\ref{sec:susceptibility_and_ergodicity_breaking} examines the static and microwave dielectric susceptibilities and the associated signatures of ergodicity breaking. Sec.~\ref{sec:vibrational_mode_analysis_and_infrared_spectrum} analyzes the polar vibrational spectrum and its contribution to the far-IR response. Sec.~\ref{sec:low_frequency_and_low_temperature_properties} investigates the excess low-frequency modes, including quasi-localized excitations and their anharmonicity. Sec.~\ref{sec:specific_heat} connects these excitations to thermal anomalies through a comparison of the specific heats of PMN and PTO. Finally, Sec.~\ref{sec:conclusions} summarizes the main findings and discusses open questions.
\begin{figure*}[t]
    \centering
    \includegraphics[width=\linewidth]{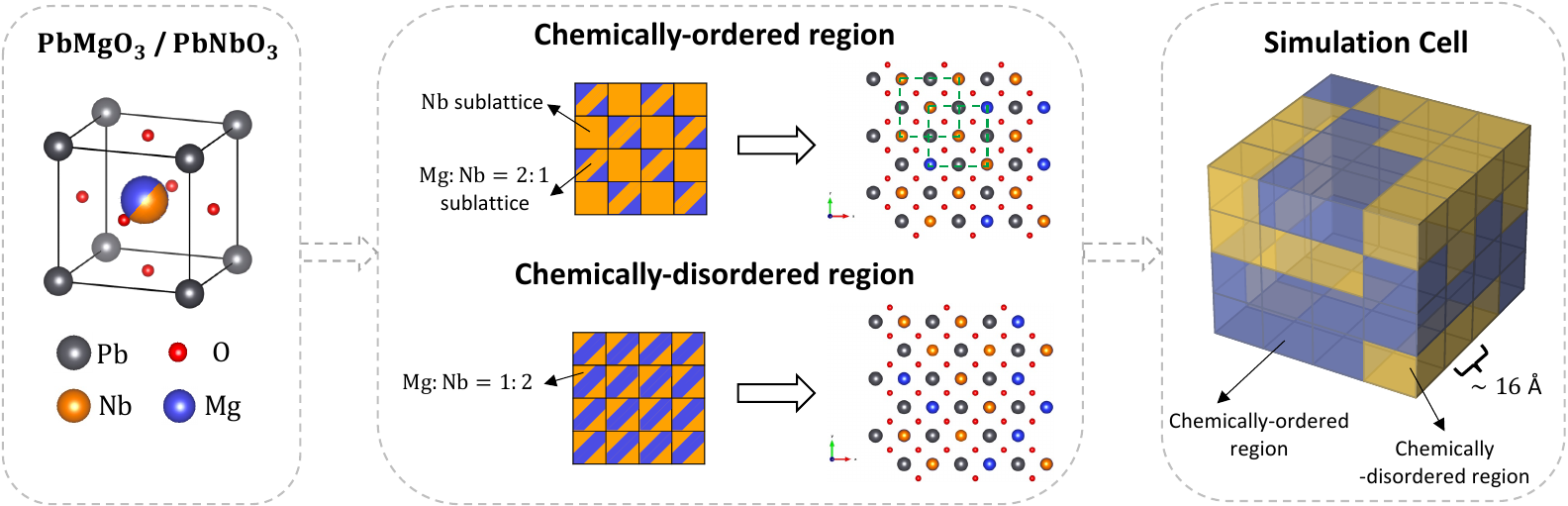}
    \caption{\label{fig:supercell} Schematic representation of the chemically-ordered regions and chemically-disordered regions. They are building blocks of the supercells adopted in MD simulations.
    }
\end{figure*}

\section{Atomistic models of PMN\label{sec:machine_learning_models}}

We follow the approach adopted by Xie \textit{et al}~\cite{pto2025prb}
who used two neural networks to study the ferroelectric phase transition in PTO. One network called Deep Potential (DP)~\cite{zhang2018deep} is used to represent the potential energy of interaction among the atoms, the other network called Deep Wannier (DW) is used to describe the polarization.

In the DP model, the potential energy of an $N$-atom configuration $\{\boldsymbol{R}_i\}_{i=1}^N$ is given by  
\begin{equation}
    E \left( \{\boldsymbol{R}_i\}_{i=1}^N \right) = \sum_i E_{\theta} \left( \{ \boldsymbol{R}_j | \boldsymbol{R}_j \in \mathcal{N}_i \} \right),
\end{equation}
where the sum extends to the $N$ atoms and $E_{\theta}$ is a scalar-valued deep neural network with learnable parameters $\theta$ that represents the contribution to the potential energy of the atoms in the neighborhood of the atom $i$.
The neighborhood is defined by a spherical cutoff radius which we take here to be equal to 6 \AA.  
The network parameters are optimized by machine learning with training on density functional theory (DFT) data for representative local configurations, covering a substantial number of configurations with different compositional disorder. We adopt the strongly constrained and appropriately normed (SCAN) approximation~\cite{sun2015strongly} of DFT in view of its good balance of precision and efficiency in applications to ferroelectric perovskites~\cite{scan2017}. 

The polarization associated with the atomic configuration $\{\boldsymbol{R}_i\}_{i=1}^N$ (up to a global gauge) is divided into the ionic contribution and the valence electron contribution~\cite{vanderbilt2018berry}. The ionic charges in the system are associated with the atomic centers at $\boldsymbol{R}_i$, each of which carries an integer ionic charge $Z_i$ equal to the total charge of the nucleus plus the corresponding core electrons. 
We use the effective point charge representation of the valence electrons provided by the Wannier centers. These are the centers of the maximally localized Wannier distributions defined by the unitary transformation of the valence orbital subspace that maximizes localization~\cite{marzari1997maximally, marzari2012maximally}. In spin saturated systems, each Wannier center carries an integer charge of -2$e$. Moreover, in PMN, each Wannier center can be uniquely associated with its nearest atom, and this correspondence is preserved along molecular dynamics trajectories. Then, it is convenient to define the Wannier centroid (WC) associated with the atom $i$ as the geometric center of the Wannier centers associated with that atom (see SM~\cite{SupplementalMaterial} for details). Its relative location with respect to its parent atom is indicated by $\boldsymbol{r}_i$ and its charge
$Q_i$ is the total charge of the parent Wannier centers.

The DW model~\cite{zhang2020deep} is trained to predict $\boldsymbol{r}_i$ using an equivariant, vector-valued neural network that takes as input the coordinates of the atoms in the neighborhood of the atom $i$. The DW model is trained on consistent SCAN-DFT data for WCs of representative configurations, and it extends beyond the conventional linear approximation commonly used to describe the valence-electron contribution to polarization.

The total polarization $\boldsymbol{P}$ of a periodic system of volume $V$ can be defined as $\boldsymbol{P} = V^{-1}\sum_i Z_i \boldsymbol{R}_i + Q_i \left( \boldsymbol{R}_i + \boldsymbol{r}_i \right)$. Polarization theory~\cite{vanderbilt2018berry} states that only changes in $\boldsymbol{P}$ modulo the polarization quantum are physically meaningful. In this paper, the change in $\boldsymbol{P}$ during the MD simulation remains within one polarization quantum, so we can take the absolute change in $\boldsymbol{P}$ as the polarization change.

The DP and DW models retain the accuracy of first-principles simulations at a much lower computational cost.
Details on DFT calculations, training, and validation of these models can be found in the Supplementary Material (SM)~\cite{SupplementalMaterial}. In addition, we employ a decomposition of the polarization into local dipole moments associated with each Mg/Nb-centered primitive cell in the SM~\cite{SupplementalMaterial}. This definition enables visualization of PNR and analysis of local dipole--dipole correlation functions~\cite{SupplementalMaterial}.

\section{Compositional and polar structures\label{sec:structural_model_and_PNRs}}

Experiments show that chemically ordered regions (COR) and chemically disordered regions (CDR) are both present at the nanoscale in PMN samples~\cite{viehland1991glassy, boulesteix1994numerical, fu2009relaxor, cabral2018gradient}.
The COR contains a double perovskite structure with two interpenetrating sub-lattices, where the B-cation sites of one sub-lattice are exclusively occupied by Nb atoms, while the B-cation sites of the other sub-lattice are randomly occupied by Mg or Nb atoms with compositional ratio 2:1. In the CDR, all B-cation sites are randomly occupied by Mg and Nb atoms with a stoichiometric ratio of 1:2.  

Recent first-principles investigations of the compositional structure of PMN in Ref.~\cite{xu2026intrinsic} clarify that COR is the dominant intrinsic compositional structure of PMN. The intrinsic compositional structure can be coherently extended during crystal growth, provided that the chemical ordering of metallic ions can be fully equilibrated. However, real-world PMN samples are prepared near melting and quenched out of equilibrium, leading to the intermixing coexistence of COR and CDR found in the samples. 

The systematic algorithm~\cite{xu2026intrinsic} for  sampling PMN's compositional disorder cannot easily scale to the large-scale compositional structures required by the scope of this work.  Here, we instead empirically model the compositional structure of PMN simulation cells, based on the experimental findings and the small-scale simulations in Ref.~\cite{xu2026intrinsic}.
We build the simulation cell as a random mixture of an equal number of COR and CDR blocks, each comprising $3\sqrt{2}\times3\sqrt{2}\times4$ primitive cells, as illustrated by Fig.~\ref{fig:supercell}. Each random mixture constitutes a distinct disorder realization.

\begin{figure*}[t]
    \centering
    \includegraphics[width=\linewidth]{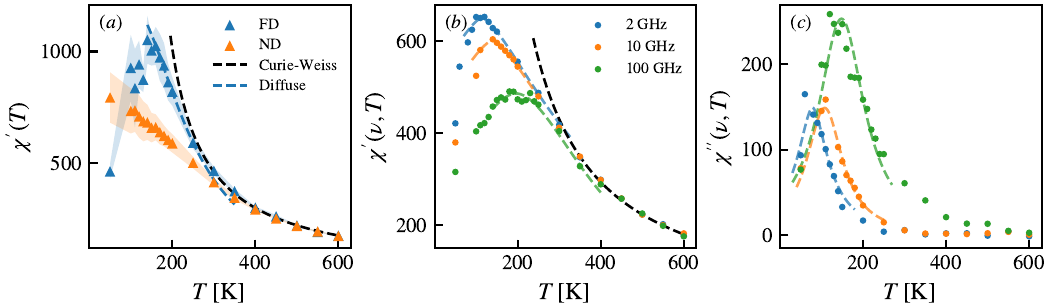}    \caption{\label{fig:dielectric_response} (a) $\chi'_{\text{FD}}$ and $\chi'_{\text{ND}}$ as functions of temperature. The shaded regions indicate the standard deviation with respect to disorder realizations. The black and blue dashed lines are fits to the Curie-Weiss and diffuse transition laws, respectively. (b) $\chi'(\nu, T)$. The black dashed line is a fit to the Curie-Weiss law. The colored dashed lines are fits to the diffuse transition law. (c) $\chi''(\nu, T)$. The colored dashed lines are guides for the eyes.  
        }
\end{figure*}

Next, we perform canonical molecular dynamics (NVT-MD) simulations on periodic supercells containing $12\sqrt{2}\times 12\sqrt{2}\times 16$ primitive cells, over the temperature interval $50$–$800$ K. We fix the lattice constants of the simulation box to the experimental values in the high-temperature paraelectric phase. The simulations are repeated for supercells with different disorder realizations to reduce stochastic biases. 
In the SM~\cite{SupplementalMaterial}, we show that our model captures the qualitative features of the atomic structure observed in experiments, such as the neutron total radial correlation function. In addition, at high temperatures, we have verified that local dipole moments are randomly oriented and continuously fluctuating. However, at approximately 250 K, soft PNR with dynamic dipole alignment begins to form in Nb-rich environments. Upon further lowering of the temperature, the PNR orientations appear frozen on the nanosecond time scale of the simulations. It is also described in the SM~\cite{SupplementalMaterial} that
nearly identical PNR are obtained by repeating the cooling procedure for the same disorder realization but starting from a different equilibrium structure at 800 K, indicating that the PNR patterns are dictated primarily by the underlying quenched compositional disorder.    

The Nb-rich character of the PNR is reflected in the local-dipole statistics. At low temperatures, Nb-centered cells develop enhanced nearest-neighbor dipolar correlations compared with Mg-centered cells. All these observations are consistent with simulations previously performed with smaller supercells in Ref.~\cite{xu2026intrinsic}, and they are consistent with the experimentally observed association between PNR and Nb-rich environments~\cite{fu2009relaxor}. 

Thus, despite the simplified representation of
compositional disorder, our model captures the essential microscopic phenomenology: a non-ferroelectric state containing heterogeneous, increasingly slow polar correlations that are pronounced in Nb-rich environments.

\section{Susceptibility and Ergodicity Breaking\label{sec:susceptibility_and_ergodicity_breaking}}

In experiments, the static dielectric susceptibility $\chi'$ of PMN exhibits different temperature-dependent behaviors when the system is cooled to low temperatures with or without a weak applied field of the order of a few kV/cm~\cite{levstik1998glassy}. In the presence of the field, the sample gradually develops a finite polarization, while in the absence of the field, the susceptibility shows a diffuse transition into a glassy nonpolar state. The two protocols differ below a characteristic temperature $T^*$ that signals ergodicity breaking~\cite{viehland1992deviation}. 

In simulations like ours, the effects of fields as small as those used in the experiments are not observable, but the loss of ergodicity can still be detected from the difference between the susceptibility predicted by the fluctuation-dissipation theorem, $\chi'_{\text{FD}}$, and that predicted by the numerical derivative of polarization with respect to an applied electric field, $\chi'_{\text{ND}}$~\cite{rabe1998first, akbarzadeh2012finite}. 
Here, 
$$\chi'_{\text{FD}}= \frac{V}{3\varepsilon_0 k_{\text{B}} T}\left[
\langle \boldsymbol{P}^2 \rangle - 
\langle \boldsymbol{P} \rangle^2  \right]_{\text{dis}},$$ 
where $\varepsilon_0$ is the dielectric permittivity of the vacuum, and 
\begin{align*}
\chi'_{\text{ND}} &= 
\frac{1}{\varepsilon_0}
\left[\lim_{\left|\boldsymbol{E}\right|\to 0}\frac{ {\langle P_{\|} \rangle}_{\boldsymbol{E}} - {\langle P_{\|} \rangle}_{0} }{\left|\boldsymbol{E}\right|}\right]_{\text{dis}} \\
&\approx 
\frac{1}{\varepsilon_0}
\left[ \frac{ {\langle P_{\|} \rangle}_{\boldsymbol{E}_0} - {\langle P_{\|} \rangle}_{0} }{\left|\boldsymbol{E}_0\right|}\right]_{\text{dis}},
\end{align*}
where $P_{\|} \equiv \boldsymbol{P} \cdot \hat{\boldsymbol{E}}$. The brackets $[\cdot ]_{\text{dis}}$ denote the average over different realizations of quenched disorder. And ${\langle P_{\|} \rangle}_{\boldsymbol{E}}$ indicates thermally averaged polarization along the applied field $\boldsymbol{E}$. In our simulations, the applied field $\boldsymbol{E}_0$ is small but observable.  We calculated $\chi'_{\text{FD}}$ with a NVT-MD simulation under zero field in which the temperature was gradually and slowly reduced from 800 to 50 K. To compute $\chi'_{\text{ND}}$, we initially performed a NVT-MD simulation under the field $E_0=\left|\boldsymbol{E}_0\right|=150$ kV/cm applied along the direction $\left[111\right]$, 
while the temperature was gradually and slowly reduced from 800 to 50 K. Then we used configurations extracted from the cooling protocol at various temperatures to start MD trajectories and calculate $\langle P_{\|} \rangle_{\boldsymbol{E}_0}$ at those temperatures. The adopted $\boldsymbol{E}_0$ was sufficiently strong to permit dielectric relaxation on the nanosecond time scale of the simulations while being sufficiently weak for the validity of linear response theory.

The calculated $\chi'_{\text{FD}}$ and $\chi'_{\text{ND}}$ are reported in Fig.~\ref{fig:dielectric_response}(a).
At low temperature, the two curves deviate from the Curie-Weiss law ($\chi'^{-1} \propto T-$116 K) and from each other. $\chi'_{\text{FD}}$ shows a broad peak around 140 K that mimics the diffuse paraelectric-relaxor transition observed in experiments at about 240 K~\cite{levstik1998glassy},
while $\chi'_{\text{ND}}$ increases monotonically with decreasing temperature, in qualitative agreement with previous simulation studies~\cite{akbarzadeh2012finite, grinberg2009molecular}. Additional statistical analysis reported in the SM~\cite{SupplementalMaterial} shows the temperature dependence of the Edwards--Anderson (EA) order parameter~\cite{edwards1975theory, binder1986spin}, which indicates that the diffuse paraelectric-relaxor transition involves a soft glass transition~\cite{akbarzadeh2012finite}.

In correspondence with the soft glass transition, both $\chi'_{\text{FD}}$ and $\chi'_{\text{ND}}$ show large statistical fluctuations for different disorder realizations. We notice that the peak amplitude of $\chi'_{\text{FD}}$ is approximately one order of magnitude lower than that measured experimentally. This discrepancy is plausibly attributable to finite-size and finite-time limitations of the simulations, which preclude access to microsecond and longer dielectric relaxation processes that likely involve rare reorientation events of PNR. These limitations, however, do not preclude resolving the dynamic response in the microwave region.

A key experimental fingerprint of relaxors, shared with spin glasses, is the characteristic temperature dependence of the complex dynamic susceptibility $\chi(\nu, T)=\chi'(\nu, T) + \mathrm{i} \chi''(\nu,T)$ for temperatures below the onset of ergodicity breaking. Experiments show that for $\nu \leq 1\ \text{GHz}$ the peak of $\chi'$ associated with the diffuse transition shifts to lower temperatures as the frequency decreases, and at the same time its amplitude increases~\cite{viehland1991glassy, levstik1998glassy, kutnjak1999slow}. The opposite behavior is shown by $\chi''$ which exhibits enhanced absorption at higher temperatures for higher frequencies~\cite{viehland1991glassy, levstik1998glassy, kutnjak1999slow}. 
The behavior of $\chi'(\nu, T)$ is described by the phenomenological law for the diffuse transition~\cite{clarke1974diffuse, bokov2000phenomenological, santos2001phenomenological, bokov2003empirical}: 
\begin{equation}
    \chi'^{-1} =  {\chi'}_{m}^{-1} \left( 1 + \left( \left( T - T_{m} \right) / \Delta_m \right)^{\gamma} \right),
    \label{eq:diffuse_dispersion}
\end{equation}
where $\chi'_m$, $T_m$, and $\Delta_m$ specify the maximum of $\chi'$, the temperature associated with the maximum, and the width of the diffuse transition peak at the given frequency $\nu$. $\gamma$ is a fitting parameter with values in the range $1 < \gamma \leq 2$ for typical relaxors~\cite{bokov2000phenomenological, santos2001phenomenological, bokov2003empirical}. For $\gamma=1$, Eq.~\eqref{eq:diffuse_dispersion} becomes the Curie-Weiss law of standard ferroelectrics such as PTO. By associating the inverse of frequency $\nu$ with a relaxation time $\tau$ ranging from 10 $\mu$s to 10 ms for frequencies in the range of 100 to 0.1 kHz, Viehland et al.~\cite{viehland1990freezing, viehland1991glassy} showed that $\tau$ depends on $T_m$ as in the Vogel-Tammann-Fulcher (VTF) law $\tau \propto \exp \left( \frac{E_a}{k_{\mathrm{B}}\left( T_f - T_m \right)} \right)$, where $E_a$ is the activation energy and $T_f$ is the freezing temperature. The VTF law also describes the relaxation dynamics in spin glasses and fragile structural glasses~\cite{vogel1921temperaturabhangigkeitsgesetz, fulcher1925analysis, tammann1926abhangigkeit}. Its super-Arrhenius character indicates that a wide continuous range of timescales diverging as $T_f$ is approached is a universal feature of glasses.         

Interestingly, the experimental behavior of $\chi(\nu, T)$ is qualitatively reproduced in our simulations when an external oscillatory field $E(t) = E_0 \sin(2\pi\nu t)$ is applied along the $[001]$ direction. Owing to the finite-time limitation of our simulations, the response function can be robustly determined, though with non-negligible statistical uncertainty, only down to the microwave regime; specifically, for frequencies $\nu = 2, 10, 100$ GHz. These frequencies exceed those typically accessed in experiments, yet they remain well below the characteristic frequency scales associated with typical lattice vibrational modes. In this study, we record non-equilibrium MD trajectories of polarization $P_z(t)$ at different temperatures $T$ under the applied external field $E(t)$, and we compute $\chi(\nu, T)=\varepsilon_0^{-1} E_0^{-1}P_z(\nu)$ from the Fourier transform of $P_z(t)$. We used trajectories that are 5 ns long for $\nu = 2$ GHz, and 1 ns long otherwise. The results are reported in
Fig.~\ref{fig:dielectric_response}(b, c). Fig.~\ref{fig:dielectric_response} also includes fits of $\chi'(T)$ and of $\chi'(\nu, T)$ using the law in Eq.~\eqref{eq:diffuse_dispersion} with $\gamma = 2$. In general, the experimental trends of $\chi'(\nu, T)$ and $\chi''(\nu, T)$
are well reproduced, the main difference being that the predicted peak values of the response are small compared to the experiments, an effect due both to the larger frequencies accessible in the simulations and to the finite size of the simulation box. The predicted temperatures at which the diffuse transition peaks occur are systematically lower, perhaps by about 100 K, than their experimental counterpart, an issue likely due to the accuracy of the adopted DFT approximation and to that of the simple model used for the quenched disorder.

\section{Vibrational-mode analysis and Infrared Spectrum\label{sec:vibrational_mode_analysis_and_infrared_spectrum}}

The low frequency behavior of $\chi(\nu, T)$ has illustrated the remarkable feature that relaxors are sensitive to an applied electric field in the microwave region. This is different from the behavior of conventional ferroelectrics such as PTO,
which are sensitive to a time-dependent electric field primarily near the frequencies of the IR active optical phonon modes. In PTO at low temperature, the IR active mode
of lowest frequency occurs at about 80 $\mathrm{cm^{-1}}$ ($\approx 2400$ GHz), and all the modes with lower frequency are acoustic modes that do not involve microscopic charge separation or couple with a macroscopic electric field. By contrast, due to compositional disorder, a relaxor such as PMN has low-frequency collective vibrational modes that exhibit microscopic charge separation. 

To better understand the origin of this phenomenology, we performed a comparative study of the vibrational spectra of PMN and PTO based on the DP model developed here for PMN and the one for PTO~\cite{pto2025prb}. Both models were trained on the same DFT functional approximation. In this study, we used smaller simulation cells than those adopted in Sec.~\ref{sec:susceptibility_and_ergodicity_breaking} to facilitate direct diagonalization of the dynamical matrix. Specifically, we used a periodic supercell containing $9 \sqrt{2} \times 9\sqrt{2} \times 12$ primitive cells for PMN and a periodic supercell containing $10 \times 10 \times 10$ primitive cells for PTO. A PMN equilibrium structure was prepared by cooling a supercell from 400 to 10 K along a NPT-MD trajectory with a cooling rate of 0.39 K/ps, followed by conjugate gradient relaxation. The resultant equilibrium structure is unique to each realization of compositional disorder and can be called an inherent structure~\cite{stillinger2015energy}. This procedure was repeated for four distinct disorder realizations. The resultant four inherent structures all have a cubic lattice constant of around $a=4.08$ \AA~, with an insignificant variance, to be compared with $a=4.04$ \AA~in the experiment~\cite{de1991structural, wakimoto2002ferroelectric}. As for PTO, the equilibrium structure is obtained with a standard protocol. The supercell is relaxed with lattice parameters fixed to the experimental low temperature values of $a=b=3.88$ \AA~ and $c/a=1.071$~\cite{mabud1979lattice}.

\begin{figure*}[t]
    \centering
    \includegraphics[width=\linewidth]{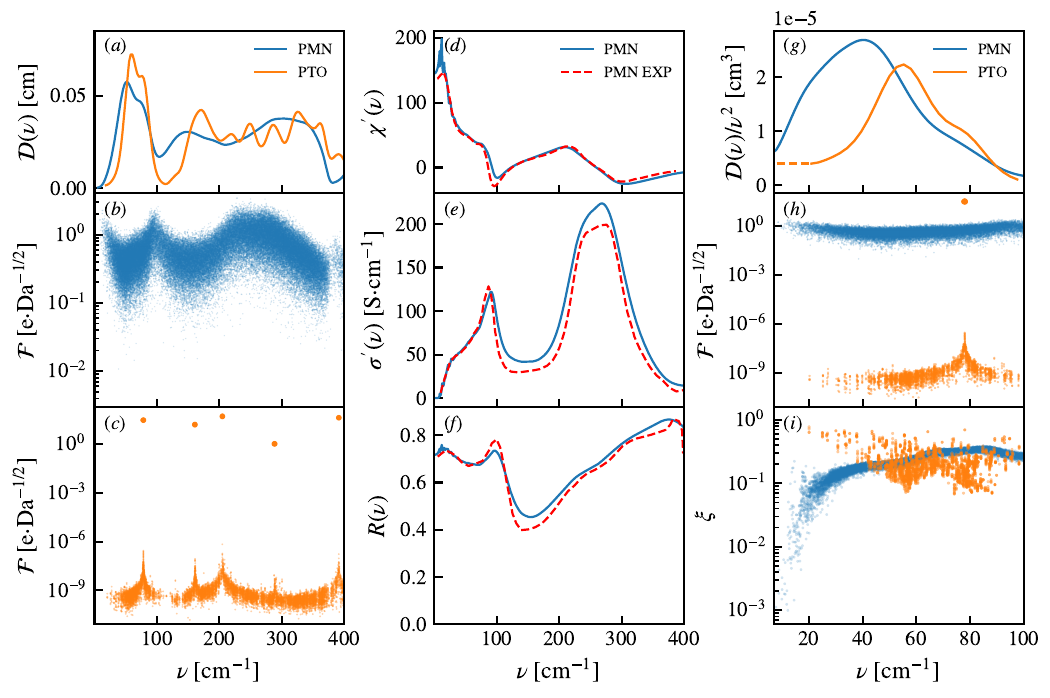}
\caption{\label{fig:PMN_PTO_ha_IR_spectra} (a) VDOS of PMN and PTO. (b) Frequency-dependent oscillator strength of vibrational modes in PMN. (c) Frequency-dependent oscillator strength of vibrational modes in PTO. Data points with $\mathcal{F}>0.1\ \mathrm{e\cdot Da^{-1/2}}$ are displayed using larger markers.
(d) Real part of the dielectric susceptibility of PMN; the noisy character of the calculated $\chi'(\nu)$ at very low frequencies is attributed to the poor statistics of the calculated modes at these frequencies. (e) Real part of the optical conductivity of PMN. (f) Reflectivity of PMN. The calculated spectra in (d), (e) and (f) are compared with the experimental data measured at $T=10$ K (red) ~\cite{PhysRevB.96.174113}. (g) Reduced VDOS in the low-frequency regime. (h) Oscillator strength of low-frequency vibrational modes in PMN and PTO. (i) Participation ratio of low-frequency vibrational modes. 
}
\end{figure*}

For the inherent structures of PMN and for the unique equilibrium structure of PTO, the dynamical matrices constrained by the acoustic sum rule~\cite{ackland1997practical} were diagonalized to compute the eigen-frequencies $\nu^{(j)}$ and the corresponding normalized eigen-modes $\tilde{\phi}^{(j)}$. 
The results in the far-IR region are reported in Fig.~\ref{fig:PMN_PTO_ha_IR_spectra}. 
The highest frequency in this region is two orders of magnitude larger than the highest frequency used in the linear response calculations discussed in Sec.~\ref{sec:susceptibility_and_ergodicity_breaking}. 

Fig.~\ref{fig:PMN_PTO_ha_IR_spectra}(a) shows the VDOS per primitive cell, denoted by $\mathcal{D}(\nu)$. 
For PMN, the VDOS has a negligible spread among the disorder realizations. This suggests minor finite size effects, and in particular, the VDOS depends mostly on the gross features of the compositional disorder. Accordingly, Fig.~\ref{fig:PMN_PTO_ha_IR_spectra}(a) reports the VDOS of PMN as an average over the four inherent structures.

While VDOS describes the frequency distribution of vibrational modes, each mode’s response to an external electric perturbation is quantified by the frequency-dependent oscillator strength, $\mathcal{F}$. For eigenmode $j$, the oscillator strength is defined via the Euclidean 2-norm as
$\mathcal{F}^{(j)}=\left\|\sum_{l=1}^{N} Z_{l}\phi^{(j)}_{l}\right\|_{2}$, where the sum is taken over all $N$ atoms of the system. Here, $\phi^{(j)}_l = m_l^{-\frac{1}{2}} \tilde{\phi}_{l}^{(j)}$ represents the mass-weighted vibrational amplitude of the $l$-th atom, where $m_l$ is its atomic mass. $Z_l$ denotes the Born effective charge tensor of the $l$-th atom. Here, $Z_l$ is assumed to depend solely on the atomic species of the $l$-th atom and is obtained from density-functional perturbation theory calculations performed for high-symmetry reference structures of PMN and PTO (see SM~\cite{SupplementalMaterial} for details). A more accurate estimate of the mode oscillator strength can be obtained by directly computing the response of polarization predicted by the DW model to an infinitesimal collective displacement along the corresponding normal-mode coordinate. We do not pursue this latter procedure because it is substantially more computationally demanding and does not alter the qualitative features of $\mathcal{F}$.

Figures~\ref{fig:PMN_PTO_ha_IR_spectra}(b) and (c) show $\mathcal{F}$ for PMN and PTO, respectively. 
The scatter plot shown in panel (b) represents the total statistics across four distinct inherent structures, each of which yields a nearly identical scatter plot.
Thus, the two orders of magnitude in the spread of the blue region reflect the fluctuation of the oscillator strength between modes with similar frequency and do not result from different realizations of disorder. Fig.~\ref{fig:PMN_PTO_ha_IR_spectra}(b) tells us that strongly coupled modes constitute a dense set in the frequency range, where the modes of maximal coupling occur at frequencies near the local minima of the VDOS, which are the regions where stronger manifestations of disorder are expected. By contrast, 
Fig.~\ref{fig:PMN_PTO_ha_IR_spectra}(c)
shows that in PTO the oscillator strength of most eigenmodes is essentially zero and only a handful of isolated modes have an oscillator strength comparable to that of the strongly coupled modes of PMN. The
frequencies and irreducible representations of the active optical modes in PTO are reported in the SM~\cite{SupplementalMaterial} and show good agreement with the experimental data of Ref.~\cite{foster1993anharmonicity}.

From the eigenvalues and eigenvectors of the dynamical matrices, we compute the complex dielectric susceptibility of PMN according to the equation: 
\begin{equation}
    \chi\left(\nu\right) = \frac{1}{\left(2 \pi\right)^2} \frac{1}{3\varepsilon_0V} \sum_{j} \frac{ \left( \mathcal{F}^{(j)} \right)^2 }{ \left(\nu^{(j)}\right)^2 - \nu^2 - \mathrm{i}\lambda^{(j)}\nu } .
    \label{eq:harmod}
\end{equation}
Here, $\mathrm{i}$ denotes the imaginary unit. We set $\lambda^{(j)}=\lambda \nu^{(j)}$, with $\lambda = 0.07$ adopted as a uniform damping factor. Using this single \textit{ad hoc} damping parameter, while the remaining parameters in Eq.~\eqref{eq:harmod} are derived from the dynamical matrix, the resulting predicted far-IR spectra show excellent agreement with the experimental measurements~\cite{PhysRevB.96.174113}, as shown in panels (d–f) of Fig.~\ref{fig:PMN_PTO_ha_IR_spectra}, which display, respectively, $\chi'(\nu)$, 
the real part of the dielectric susceptibility, $\sigma'(\nu)=2\pi \nu \varepsilon_0 \chi''\left(\nu\right)$, the real part of the optical conductivity, and $R(\nu)$, the optical reflectivity. $R\left(\nu\right) = \left| \frac{\sqrt{\varepsilon_{\infty} + \chi(\nu)} - 1}{\sqrt{\varepsilon_{\infty} + \chi(\nu)} + 1} \right|^2$ is fully determined by $\chi'(\nu)$ and $\sigma'(\nu)$, along with the high-frequency dielectric constant $\varepsilon_{\infty}$.

The close agreement between simulations and experiments enables assigning far-IR spectral features to the underlying microscopic vibrational modes. In particular, 
the two broad peaks of $\sigma'$ located at approximately 80 $\mathrm{cm^{-1}}$ and 260 $\mathrm{cm^{-1}}$ should be attributed, respectively, to the broad features of the oscillator-strength spectrum in Fig.~\ref{fig:PMN_PTO_ha_IR_spectra}(b)
around the same frequencies.  In contrast, the spectra of PTO in the same frequency interval (see SM~\cite{SupplementalMaterial}) are characterized by substantially sharper, narrower features, attributable to the set of dominant polar modes highlighted in Fig.~\ref{fig:PMN_PTO_ha_IR_spectra}(c). 

These findings underscore the critical contribution of disorder to the distinctive spectral characteristics of the PMN relaxor in the far-IR regime.

\section{Low-frequency vibrational mode anomalies\label{sec:low_frequency_and_low_temperature_properties}}

Fundamental difference between PMN relaxor and conventional ferroelectrics also lies in the low-frequency part of the VDOS. Fig.~\ref{fig:PMN_PTO_ha_IR_spectra}(g) plots the reduced VDOS, $\mathcal{D}(\nu)/\nu^2$, as a function of $\nu$ for $\nu \leq 100~\mathrm{cm}^{-1}$. PMN shows a broad excess of low-frequency modes beyond the Debye model, including both non-phononic and phonon-like modes. 
In contrast, PTO follows the Debye model at sufficiently low frequency because acoustic phonons in PTO obey a linear dispersion when $\nu \leq 20~ \mathrm{cm}^{-1}$, which has been established through inelastic neutron scattering~\cite{tomeno2006lattice}. Although the finite supercell of PTO used in our real-space vibrational analysis does not directly resolve these long-wavelength acoustic phonons, the trend of the reduced VDOS approaching $\nu = 20~ \mathrm{cm}^{-1}$ suggests a flat plateau for $\nu \leq 20~ \mathrm{cm}^{-1}$ (dashed orange line in Fig.~\ref{fig:PMN_PTO_ha_IR_spectra}(g)), a result consistent with the scattering experiments~\cite{tomeno2006lattice}.

The broad and prominent peak in the reduced VDOS of PMN is typically called a boson peak~\cite{mizuno2017continuum}.  The broad boson peak is fundamentally different from the narrow peak in the reduced VDOS of PTO. The latter results from the non-linear dispersion of acoustic phonons near the zone boundary~\footnote{frequencies of zone boundary transverse acoustic phonons in PTO are distributed around 55 $\mathrm{cm}^{-1}$~\cite{tomeno2006lattice}}, which is standard behavior of an ordered crystal. 

For $\nu \leq 100~\mathrm{cm}^{-1}$, Fig.~\ref{fig:PMN_PTO_ha_IR_spectra}(h) directly contrasts the oscillator-strength distributions of PMN and PTO, which are shown separately in Fig.~\ref{fig:PMN_PTO_ha_IR_spectra}(b) and (c).
Here, all modes of PMN are strongly coupled with the electric perturbation down to the lowest frequencies in the far-IR region, while the lowest-frequency active IR mode in PTO has $\nu \approx 80$ $\mathrm{cm}^{-1}$ and is an isolated mode. This further corroborates the conclusion that compositional disorder mediates coupling between vibrational modes and electric polarization across a broad frequency range. The presence of active modes down to the lowest frequencies also explains the increase in the dielectric susceptibility of PMN at low frequencies in Fig.~\ref{fig:PMN_PTO_ha_IR_spectra}(d). 

\begin{figure}[bt]
    \centering
    \includegraphics[width=0.9\linewidth]{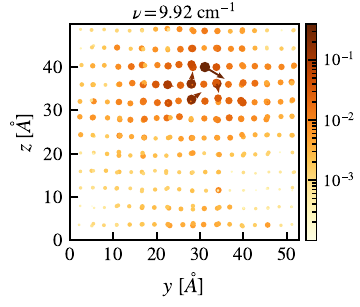}
    \caption{A representative localized eigenmode ($\nu=9.92~\mathrm{cm}^{-1}$) from an arbitrary inherent structure of PMN. Pb atoms within a thin layer parallel to the \(yz\) plane are extracted from the simulation cell and displayed at their $(y,z)$ coordinates. The color and size of each circle encode the magnitude of the normalized eigenvector amplitude. Arrows indicate the direction and magnitude of its in-plane components; for clarity, arrows are shown only for sites with a large amplitude. 
    }\label{fig:QLE} 
\end{figure}

Moreover, we find that the lowest-frequency modes in PMN are localized, in sharp contrast to the extended modes in PTO.  The fraction of atomic volume shared by the $j$-th vibrational eigenmode can be measured by the participation ratio 
\begin{equation}
\xi^{(j)} = \frac{\left( \sum_{l=1}^N \|\phi^{(j)}_l\|_2^2 \right)^2}{N \sum_{l=1}^N \|\phi^{(j)}_l\|_2^4},
\end{equation}
which is reported for PMN and PTO in Fig.~\ref{fig:PMN_PTO_ha_IR_spectra}(i). It shows that, while the modes of PTO have localization consistent with spatially extended phonon modes, in PMN, only the higher frequency modes in the figure satisfy this criterion. When $\nu$ is about $40~\mathrm{cm}^{-1}$ or lower, the PMN modes are progressively more localized than the modes of similar frequency in PTO. Especially for $\nu \leq 20 ~\mathrm{cm}^{-1}$,  the participation ratio of PMN drops dramatically by about two orders of magnitude, indicating the presence of non-phononic modes in which only a relatively small group of atoms moves substantially. This picture is confirmed by the visualization of a representative eigenmode of PMN in Fig.~\ref{fig:QLE}, which shows that while the core atoms (larger dots with arrows indicating displacements) exhibit large displacements, the displacements of the remaining atoms are two orders of magnitude smaller and decay with the distance from the core. 
We observe that the largest atomic displacements are generally associated with Pb atoms. A representative core of a localized mode comprises a few dozen Pb atoms and exhibits an estimated spatial extent of approximately 1 nm.

\begin{figure}[tb]
    \centering
    \includegraphics[width=\linewidth]{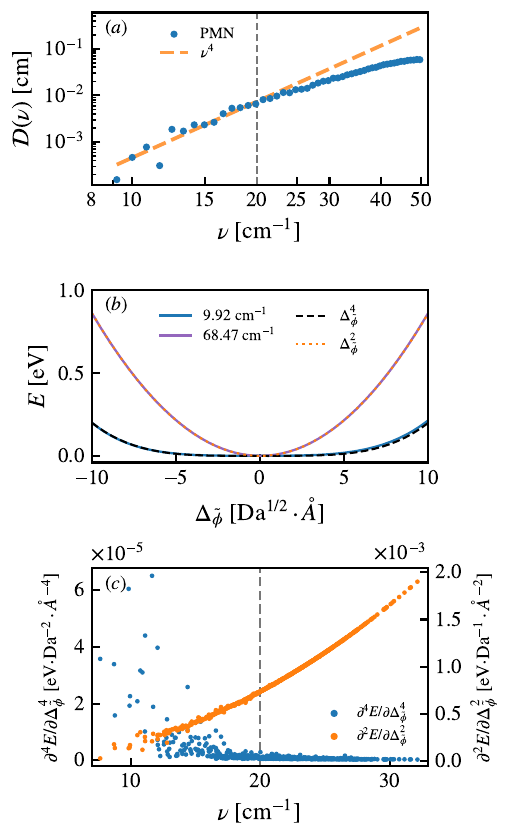}
\caption{
(a) Low-frequency asymptotic behavior of the VDOS in PMN. (b) Potential energy surfaces were computed by displacing atoms along selected vibrational modes. The extended mode ($\nu=68.47\mathrm{cm}^{-1}$) shows a quadratic dependence, whereas the QLE mode ($\nu=9.92\mathrm{cm}^{-1}$) is anharmonic. The dashed black and dotted orange lines denote the best quartic and quadratic fits, respectively.
(c) The strengths of the harmonicity and the anharmonicity of PES along vibrational modes. 
}\label{fig:harmonic_analysis_dos_cv} 
\end{figure}

These localized modes closely resemble the quasi-localized excitations (QLE) found in model structural glasses and identified therein as non-phononic vibrations localized by disorder, whose amplitude decays at a large distance from the core as the inverse of the squared distance due to elastic coupling with the more distant atoms~\cite{lerner2021low}. Studies of model structural glasses~\cite{laird1991localized, lerner2016statistics, kapteijns2018universal, lerner2021low} and of spin glasses~\cite{baity2015soft} established that the QLE contribution to VDOS scales like $\mathcal{D}(\nu)\propto\nu^4$, independent of the space dimensionality. By contrast, the phonon contribution to VDOS at low frequency depends on the dimensionality and scales in 3D like $\mathcal{D}(\nu)\propto\nu^2$. In Fig.~\ref{fig:harmonic_analysis_dos_cv}(a), we report the calculated VDOS of PMN up to $\nu=50~\mathrm{cm}^{-1}$ using a double logarithmic scale. The figure indicates that the VDOS of PMN indeed exhibits a low-frequency scaling behavior consistent with $\mathcal{D}(\nu) \propto \nu^4$ when $\nu < 20\ \mathrm{cm}^{-1}$. For $\nu$ increases above 20 $\mathrm{cm}^{-1}$, the VDOS undergoes a crossover to scaling regimes characterized by smaller exponents. The vibrational modes within this crossover region remain strongly influenced by disorder and therefore cannot be classified as conventional acoustic phonons.

An interesting question is how the QLE cores are spatially distributed. Their radial distribution function is reported as a histogram in the SM~\cite{SupplementalMaterial}. The poor resolution reflects the limited statistics, but some tendency to cluster at relatively short distances is discernible. A coordination number evaluated at a radial distance of 10~\AA{} indicates that, on average, each QLE core has approximately four to five neighboring QLE cores within this local proximity. However, at present, this remains a speculative remark. A more conclusive answer to this issue will come from future studies with better statistics.

Beyond their spatial localization, QLE are also distinguished by a strongly anharmonic energy landscape: the potential energy along a QLE eigenvector is typically dominated by a quartic term, whereas that along a non-QLE eigenvector is predominantly quadratic. This contrast is illustrated in Fig.~\ref{fig:harmonic_analysis_dos_cv}(b) for representative QLE and phonon modes. In particular, for extended phonon-like modes, the potential energy has almost perfect quadratic behavior up to relatively large amplitudes. By contrast, for QLE, the quadratic approximation to potential energy holds only for very small amplitudes ($\Delta_{\tilde{\phi}}\leq 0.5\  \mathrm{Da^{1/2}\mathrm{\AA}}$), while for larger amplitudes, the scaling is quartic, consistent with the findings in model structural glasses~\cite{lerner2016statistics, lerner2021low}. This signals a strong incipient anharmonicity~\cite{bouchbinder2021low}, whose emergence at low frequency is illustrated in Fig.~\ref{fig:harmonic_analysis_dos_cv}(c) by the behavior of the second and fourth derivatives of the potential energy with respect to the mode amplitude. The quartic anharmonicity, absent at high frequency, appears below $\nu = 20$ $\mathrm{cm}^{-1}$ and become more pronounced with decreasing frequency.

\section{Thermal anomalies\label{sec:specific_heat}}

The distinctive microscopic characteristics of the anomalous low-frequency modes examined in the preceding section manifest measurably in the thermodynamic response of PMN. 
A phenomenological low-temperature description of specific heat in insulating glasses can be expressed as $C(T)=aT + C_{\text{vib}}(T)$.
The term $aT$ is usually attributed to effective quantum tunneling modes in phenomenological theories of glasses~\cite{anderson1972anomalous, binder2011glassy}, which are beyond the description of standard harmonic theories of vibrational excitations.  $C_{\text{vib}}(T)$ represents the contribution of vibrational modes. 
In the Debye model, $C_{\text{vib}}(T)=bT^3$, which ignores deviations due to non-linear dispersion and disorder-induced effects.

\begin{figure}[t]
    \centering
    \includegraphics[width=\linewidth]{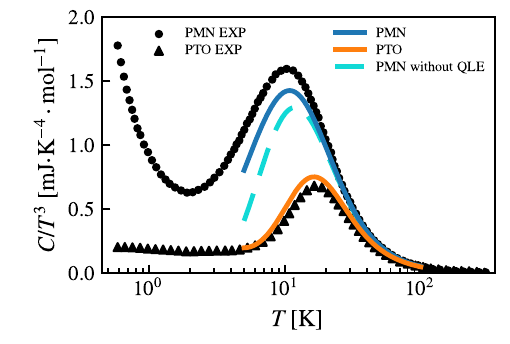}
\caption{\label{fig:harmonic_analysis_specific_heat} The reduced specific heat, $C/T^3$, peaks at approximately 10 K for PMN with an enhanced magnitude compared to that for PTO.}
\end{figure}

The experimental reduced specific heat of PMN and PTO~\cite{tachibana2009thermal} is plotted in Fig.~\ref{fig:harmonic_analysis_specific_heat} using solid black markers. In PTO, no appreciable linear-in-$T$ contribution is observed, and $C/T^3$ approaches a plateau as $T \rightarrow 0$. The maximum of the reduced specific heat appears at approximately 16 K due to a deviation from the linear acoustic-phonon dispersion assumed by the Debye model.  
In PMN, there is a substantial linear-in-$T$ contribution to specific heat, manifested as a divergent $a/T^2$ term as $T \rightarrow 0$. The reduced specific heat also exhibits a maximum centered at approximately 10 K, with a more pronounced magnitude than that of PTO. This reflects the broad disorder-induced excess of both QLE and delocalized phonon-like modes in PMN, whereas the maximum in PTO originates from ordinary acoustic-phonon dispersion.

We can validate these experimental results by calculating the vibrational contribution to specific heat through the formula $C_{\text{vib}} = \int \mathcal{D}\left(\nu\right) \mathcal{C}\left(\nu, T\right) d\nu$, where 
$\mathcal{C}\left(\nu, T\right)$ is the contribution of the mode of frequency $\nu$ at temperature $T$. 
For PTO, $\mathcal{C}\left(\nu, T\right)$ is given by the standard expression for quantum harmonic oscillators, which is also valid for the phonon-like modes of PMN; however, the effect of anharmonicity should be taken into account when dealing with the QLE. This is achieved by numerically solving the Schr\"{o}dinger equation for the one-dimensional effective Hamiltonian of the mode under a potential like the one shown in Fig.~\ref{fig:harmonic_analysis_dos_cv}(b). 

The calculated reduced specific heats $C_{\text{vib}}/T^3$ of PMN and PTO agree well with the experimental data for $T>5$ K in Fig.~\ref{fig:harmonic_analysis_specific_heat}. We can access only temperatures higher than 5 K due to the finite size of the simulation box and the missing anharmonic interactions in PMN. The computed reduced specific heat of PMN shows a maximum at approximately 10.8 K with a magnitude of 1.43 $\mathrm{mJ\cdot K^{-4}\cdot mol^{-1}}$. In PTO, the maximum occurs at approximately 16.4 K and has a magnitude of 0.75 $\mathrm{mJ\cdot K^{-4}\cdot mol^{-1}}$. The shift to lower temperatures and the enhancement of the maximum in PMN are common features associated with anomalous low-frequency modes that distinguish disordered insulating systems from their ordered counterparts~\cite{tachibana2009thermal}. Such shifts and enhancements are sometimes also called a ``boson peak'' in the literature. To avoid ambiguity, we only adopt the microscopic definition of ``boson peak'' based on the VDOS.  
Despite our simplified disorder model, our calculated maximum of reduced specific heat recovers approximately $89 \%$ of the experimental value, with a considerable contribution from the QLE. More contributions come from the main body of the prominent boson peak of PMN, which contains many vibrational modes with intermediate spatial localization character. This can be seen from the curve in Fig.~\ref{fig:harmonic_analysis_specific_heat} labeled ``PMN without QLE,'' which shows that QLE account for only part of the excess vibrational contribution. 
This result highlights the critical role of the compositional disorder in PMN that distinguishes it chemically from the ordered PTO. The compositional disorder results in anomalous low-frequency modes and further leads to thermodynamic anomalies in PMN. 

Not addressed in the present discussion is the experimentally observed divergent contribution to $C/T^3$ in PMN for temperatures below 1 K, a behavior that is widely regarded as a universal characteristic of insulating glasses.   This is beyond the scope of our vibrational analysis and would require a non-perturbative approach such as a path-integral treatment of quantum statistical mechanics.

\section{Conclusions\label{sec:conclusions}}

We studied the PMN relaxor with molecular dynamics simulations using deep neural networks trained on first-principles electronic structure data to accurately model the interatomic interactions and polarization. We adopted a heterogeneous model of disorder in which COR coexist with CDR.

The simulations qualitatively reproduced the experimental results on the temperature behavior of $\chi(\nu, T)$ at low frequency, showing a loss of ergodicity and glassy relaxation manifested by the frequency-dependent shifts of the peaks of dielectric susceptibility. However, due to the finite size of our simulations in space and time, we could only explore the response of the system to perturbations with frequencies in the microwave region rather than in the kilohertz and sub-kilohertz regimes, largely missing the exponential growth of the relaxation times observed in experiments as we approached the freezing temperature. This, along with the limits of our model of quenched disorder missing long range fluctuations~\cite{xu2026intrinsic}, is likely the reason why the calculated susceptibility at the peak of the diffuse transition is at least an order of magnitude smaller than in experiments, a finding common to atomistic simulation studies of relaxors~\cite{grinberg2009molecular, akbarzadeh2012finite}. The challenge of simulating long time relaxation dynamics in computational models of relaxors is similar to the challenge of simulating realistic cooling rates in computer models of structural glasses, as long relaxation times are not accessible on the time and size scales of molecular dynamics simulations.  When glassy dynamics is absent, as in ferroelectric PTO, atomistic simulations have no problem predicting the magnitude of the Curie-Weiss peak~\cite{pto2025prb}.

In the simulations, a manifestation of the glassy dynamics was the formation at low temperatures of PNR with polarization directions frozen on the nanosecond time scale. PNR are also observed in experiments, but there they undergo relaxations on much longer time scales. In both experiments and simulations, the PNR are associated with Nb rich environments, a finding we rationalize with the fact that neighboring Nb cells have stronger dipolar correlations than neighboring cells involving one or two Mg atoms. 

A distinctive feature of our work
was the study of the vibrational modes of PMN for a representative set of inherent structures uniquely determined by realizations of compositional disorder. From the diagonalization of the corresponding dynamical matrices, we calculated the complex dielectric susceptibility $\chi(\nu)$ in the harmonic approximation, finding a negligible dependence on disorder realizations and excellent agreement with the experimental far-IR spectra at $T=10$ K. On the one hand, this implies that the inherent structures of our model are realistic, despite the limitations in the description of glassy relaxation. On the other hand, it suggests that different inherent structures share similar features and that the fluctuations in disorder do not affect the far-IR response.
     
A main finding of our vibrational analysis is that the compositional disorder, which produces microscopic charge separation, polarizes all vibrational excitations, so that a continuum of polar modes covers the entire spectrum. This is in stark contrast to a conventional ferroelectric like PTO in which only a handful of optical modes are IR active. The continuous mode response suggests a continuous distribution of relaxation times ranging from picoseconds to nanoseconds in our simulations.
It is plausible that such a multiscale relaxation mechanism amplified by thermal anharmonicity could yield the glassy relaxation dynamics observed across the diffuse transition. The other major effect of disorder is the excess of low-frequency vibrational modes, resulting in the so called boson peak in the reduced VDOS. Among these modes, those of lowest frequency are not phonon-like but show large displacements of isolated atomic groups and small displacements of the surrounding atoms, as if a high strain spot were coupled to an elastic continuum.  Disorder-induced modes having similar characteristics were called QLE in the context of structural glass models, where their VDOS was found to follow the universal scaling law $\mathcal{D}\left(\nu\right) \propto \nu^4$ independently of the dimensionality and form of the interatomic interaction ~\cite{bouchbinder2021low}. We find that quartic scaling with frequency is also obeyed by the QLE in PMN, further supporting the universality of this disorder feature. In addition to structural glass models, QLE have also been predicted for spin glass Hamiltonians~\cite{baity2015soft}. Our study is the first report of the occurrence of QLE in a realistic model of relaxors derived from first-principles quantum mechanical theory.

The anomalous low-frequency modes lead to thermal anomalies. Our calculations predict specific heats of PTO and PMN for $T>5$ K. The predicted intensity and location of the maximum in the reduced specific heat are in close agreement with experimental observations.

The QLE in PMN and, generally, in model disordered systems~\cite{laird1991localized, baity2015soft, lerner2016statistics, kapteijns2018universal, lerner2021low}, have been detected via harmonic analysis. However, they exhibit larger incipient anharmonicity with lower frequency, suggesting that they may be precursors of strongly anharmonic excitations dominant in the ultralow-frequency regime. Such excitations would explain the anomalous thermal properties of insulating glasses below 1 K. A popular phenomenological model proposes that these excitations are due to localized groups of atoms forming quantum two-level systems (TLS) with a wide distribution of energy-level splittings and a nearly constant VDOS to produce a linear-in-$T$ contribution to specific heat~\cite{anderson1972anomalous, phillips1981amorphous, binder2011glassy}. The possible existence of TLS and their connection to QLE was explored in several model systems~\cite{baity2015soft, mocanu2023microscopic}, but a clear atomistic identification of TLS as the origin of thermal anomalies is still lacking. Alternative models suggest that the long range elastic interaction ($\sim r^{-3}$) between QLE may be sufficient to explain the anomalous specific heat~\cite{baranovskii1980elementary, yu1988low, clare1989interacting}. These issues lie beyond the scope of this work and are left for future investigation.

\textit{Data availability}---The DP model, DW model, and their training parameters that support the findings of this study will be publicly available at Zenodo.

\textit{Acknowledgements}---We thank Karin M. Rabe, Yixiao Chen, Xinyu Xu, and Shi Liu for fruitful discussions. This work was mainly conducted within the Computational Chemical Science Center: Chemistry in Solution and at Interfaces funded by the U.S. Department of Energy under Award No. DE-SC0019394. K.C. and R.C. also acknowledge funding from the U.S. Department of Energy under Award No. DE-SC0026745.
P.X. was also supported by the Alvarez Fellowship of Lawrence Berkeley National Lab. The authors are pleased to acknowledge that the work reported in this paper was performed using the Princeton Research Computing resources at Princeton University. This research also used resources of the National Energy Research Scientific Computing Center (NERSC) operated under Contract No. DE-AC02-05CH11231 using NERSC award ERCAP0021510 and AI4Sci@NERSC Award No. DDR-ERCAP 34636.

\bibliography{ref}
 
\end{document}


\title{Supplemental Material for ``Polar vibrational excitations and glass-like response in relaxor PMN''}

\author{Kehan Cai}
\affiliation{%
 Department of Chemistry, Princeton University, New Jersey 08544, U.S.A
}%

\author{Pinchen Xie}
\affiliation{
 Program in Applied and Computational Mathematics, Princeton University, New Jersey 08544, U.S.A
}
\affiliation{Applied Mathematics and Computational Research Division, Lawrence Berkeley National Laboratory, Berkeley, California 94720, USA}

\author{Yifan Li}
\affiliation{%
 Department of Chemistry, Princeton University, New Jersey 08544, U.S.A
}%

\author{Roberto Car}
\affiliation{%
 Department of Chemistry, Princeton University, New Jersey 08544, U.S.A
}%
\affiliation{
 Program in Applied and Computational Mathematics, Princeton University, New Jersey 08544, U.S.A
}%
\affiliation{%
 Department of Physics, Princeton University, New Jersey 08544, U.S.A
}%
\affiliation{
 Princeton Materials Institute, Princeton University, New Jersey 08544, U.S.A
}%

\maketitle


\section{\label{sec:SCAN-DP}SCAN-based Deep Potential}

We base our study of PMN on density functional theory with SCAN (strongly constrained and appropriately normed) approximation~\cite{sun2015strongly}. We use VASP~\cite{kresse1996efficient, kresse1996efficiency} and the projector augmented wave (PAW) potentials~\cite{kresse1999ultrasoft} for all DFT calculations. 
The semicore $2s$-states are taken as valence states for O. The semicore $2p$-states are taken as valence states for Mg. $5d$, $6s$, and $6p$ states are taken as valence states for Pb atoms. $4p$, $5s$, and $4d$ states are taken as valence states for Nb atoms. The energy cutoff of the plane-wave basis is set as 403.93 eV. The smallest allowed spacing between k-points is 0.6 $\mathrm{\AA^{-1}}$. Gamma point is always included in the k-points.

\begin{figure}[b]
    \centering
    \includegraphics[width=0.6\linewidth]{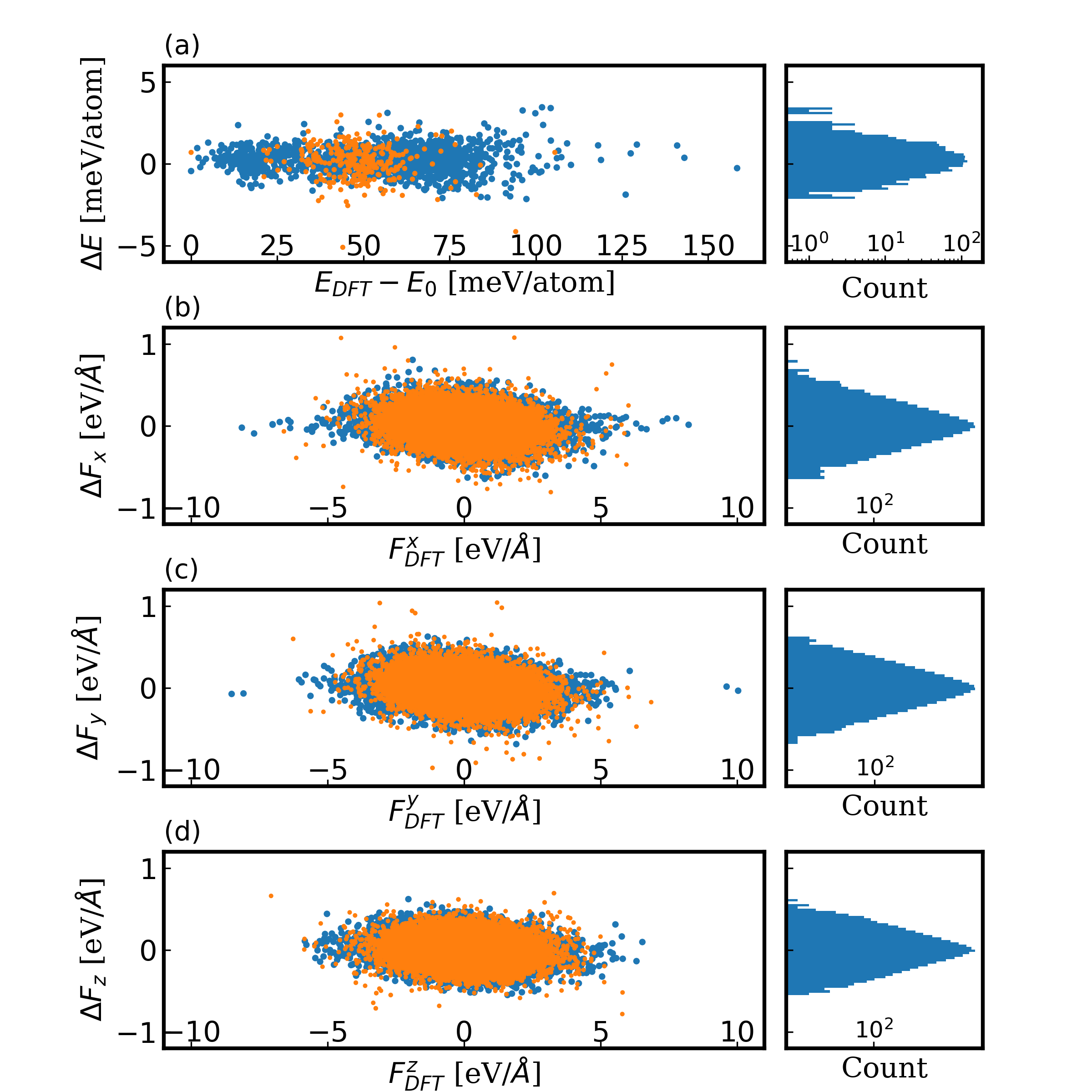}
    \caption{Error distributions of the DP model on the training set (blue) and testing set (orange).} 
    \label{Fig:dperror}
\end{figure}

We train a Deep Potential model for PMN with DeePMD-Kit~\cite{zeng2023deepmd} based on SCAN-DFT. The atomic configurations in the dataset are collected with the active learning framework DPGEN in the pressure interval $[1,3\times 10^4]$ bar and the temperature interval $[150, 650]$ K. The active exploration of configurations spaces is done with NPT-MD simulations starting from initial PMN configurations of different sizes and different disordering patterns. 
Specifically, there are three types of initial configurations: (1) Supercells of 36 PMN units (180 atoms) within the charge-balanced ``random-site" description for $\mathrm{A(B'_{1/3}B''_{2/3})O_3}$-type system: one of the B-sublattices is occupied only by $\mathrm{Nb^{5+}}$ ions while the other one contains a random distribution of $\mathrm{Nb^{5+}}$ and $\mathrm{Mg^{2+}}$ ions in a 2:1 ratio~\cite{davies2000chemical}; (2) Supercells of 72 PMN units within the charge-balanced ``random-site" description; (3) Supercells of 27 PMN units where $\mathrm{Nb^{5+}}$ and $\mathrm{Mg^{2+}}$ ions are randomly distributed on the B-lattice in a 2:1 ratio. 

The active learning procedure consists of 15 iterations of NPT-MD explorations. We use the data from 12 iterations, together with the initial dataset manually generated before active learning, as the training set, which contains DFT energy and force labels for 1386 different atomic configurations. The data from the other 3 iterations are used as the testing set, containing 300 different atomic configurations. With the training set, we obtained a DP model with a short-range cutoff $6$ \AA. The error distributions on the training set and testing set are given by Supplementary Fig.~\ref{Fig:dperror}. We report a root-mean-squared error (RMSE) of 0.8 meV/atom and 1.0 meV/atom on the training set and testing set, respectively.

\section{\label{sec:SCAN-DW}SCAN-based Deep Wannier}

For the SCAN-DFT data collected within the first 7 iterations of the active learning procedure (700 configurations in total), we consistently calculated the maximally localized Wannier functions (MLWF) for the valence electrons with VASP and the Wannier90 code~\cite{mostofi2014updated}. The geometric center of each MLWF is called its Wannier center. 

For all these data, we found for each type of atom there is a fixed number of Wannier centers localized around it. For each Pb atom, we can unambiguously associate it with 6 Wannier centers localized around it. Similarly, each Mg atom can be associated with 3 local Wannier centers, each O atom with 4 local Wannier centers, and each Nb atom with 3 local Wannier centers. This allows us to define for each atom its corresponding  Wannier centroid located at the geometric center of all its associated Wannier centers. The effective charge of each Wannier centroid is then the sum of the effective charge (-2e) of its constitutive Wannier centers. For example, the Wannier centroid associated with each Pb atom has an effective charge -12e. 

Let $\boldsymbol{r}_i$ denote the displacement, i.e. off-centering, of a Wannier centroid, with effective charge $Q_i$, from its parent atom $i$ located at $\boldsymbol{R}_i$. The total polarization $\boldsymbol{P}$ of a periodic system of volume $V$ can be defined as $\boldsymbol{P} = V^{-1}\sum_i Z_i \boldsymbol{R}_i + {Q}_i \left( \boldsymbol{R}_i + \boldsymbol{r}_i \right)$. The theory of polarization~\cite{resta2007theory} prescribes that, in a periodic system, the polarization is defined modulo a polarization quantum. Hence, only the change of $\boldsymbol{P}$ modulo the polarization quantum has physical significance. In this paper, the change of $\boldsymbol{P}$ over a MD simulation does not exceed the polarization quantum. So we will simply refer to the change in $\boldsymbol{P}$ as the change of polarization. In addition, we call $\boldsymbol{P}_{\mathrm{ion}} = V^{-1}\sum_i \left(Z_i + {Q}_i\right) \boldsymbol{R}_i$ and $\boldsymbol{P}_{\mathrm{ele}} = \boldsymbol{P}-\boldsymbol{P}_{\mathrm{ion}} = V^{-1}\sum_i {Q}_i {\boldsymbol{r}}_i$ respectively the ionic and electronic contribution to the polarization. While $\boldsymbol{P}_{\mathrm{ion}}$ is a multi-linear function of atomic positions, $\boldsymbol{P}_{\mathrm{ele}}$ is not. To capture the non-linearity, we can train a neural network model of ${\boldsymbol{r}}_i = \boldsymbol{W}_{\theta_W} \left( \{ \boldsymbol{R}_j | \boldsymbol{R}_j \in \mathcal{N}_i \} \right)$, based on the SCAN-DFT data, taking the positions of atoms $\boldsymbol{R}_j$ in the neighborhood $\mathcal{N}_i$ of atom $i$ as inputs. We call such a model the Deep Wannier model. The Deep Wannier model is short-ranged, based on the assumption that MLWFs associated with occupied orbitals in an insulating ground state depend mainly on local chemical environment.

\begin{figure*}[b]
      \centering
      \includegraphics[height=0.32\textwidth]{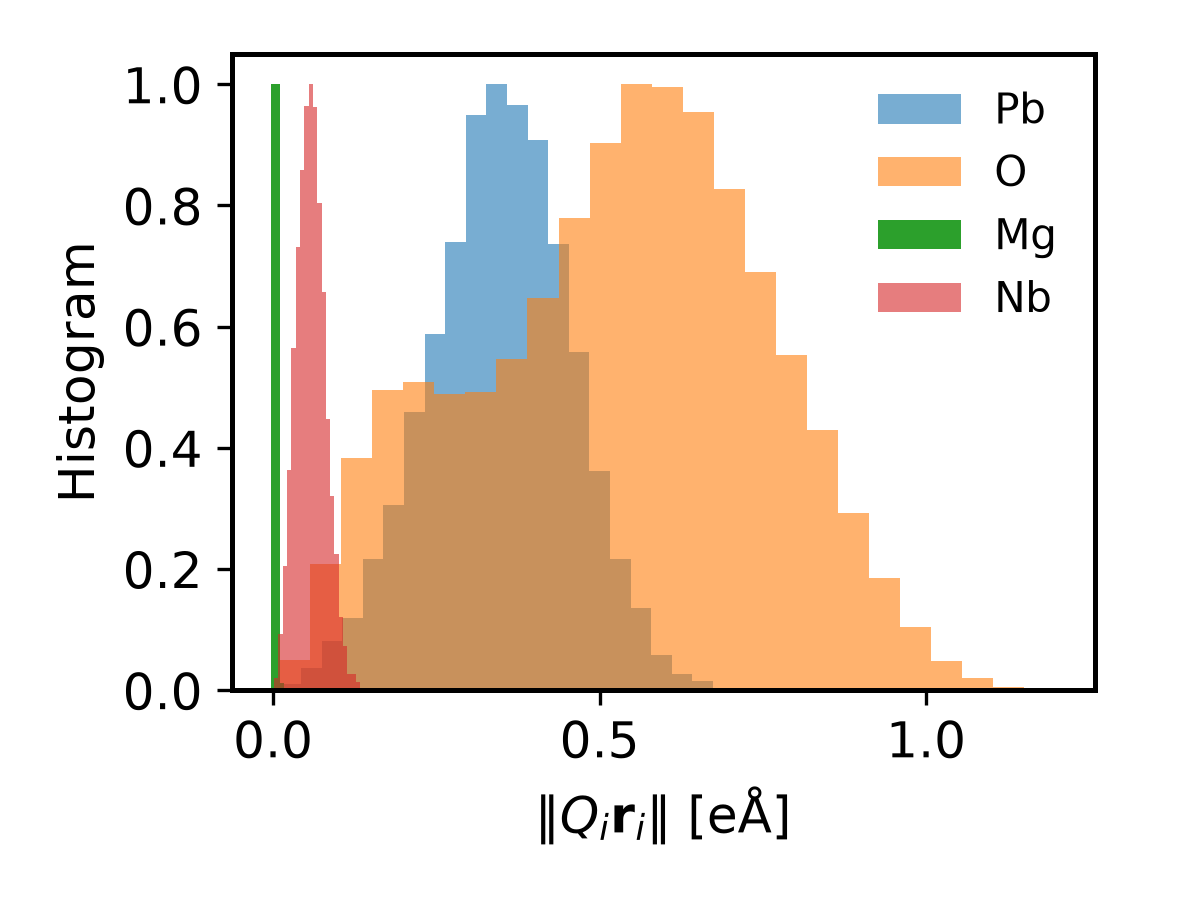}
      \includegraphics[width=0.42\textwidth]{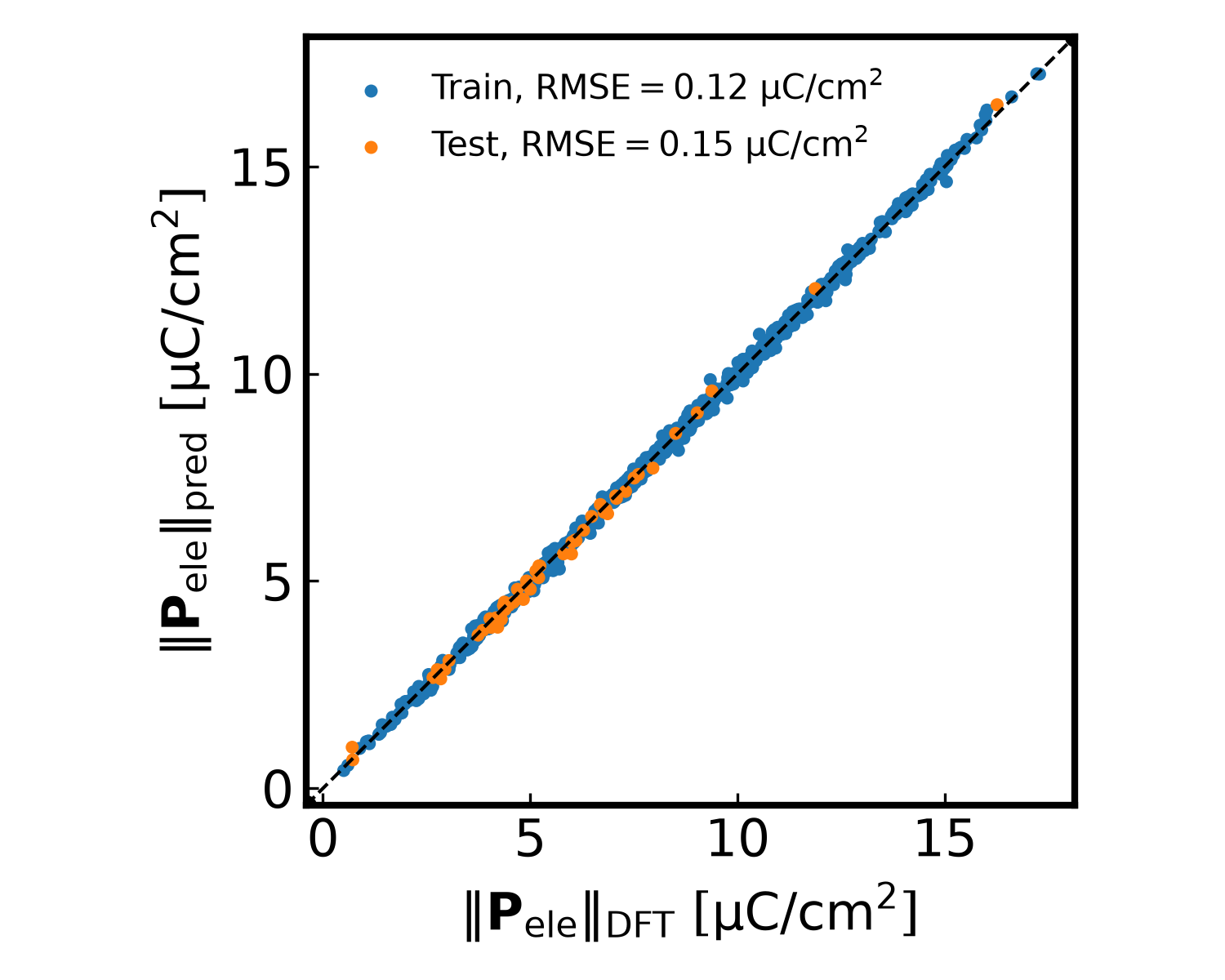}
      \caption{\label{Fig:wannier_dist}
      (Left) Normalized histograms of $\|{Q}_i {\boldsymbol{r}}_i\|$.
      (Right)  Parity plot comparing $\|\mathbf{P}_{\text{ele}}\|$ predicted by the Deep Wannier model with corresponding DFT reference values for the training set and test set.
      }
  \end{figure*}

In principle, the determination of $\boldsymbol{P}_{\mathrm{ele}}$ requires ${\boldsymbol{r}}_i$ of all atoms. However, not all types of atoms contribute significantly to $\boldsymbol{P}_{\mathrm{ele}}$. Supplementary Fig.~\ref{Fig:wannier_dist}~(Left) plots the normalized probability distribution of $\|{Q}_i {\boldsymbol{r}}_i\|$ in our dataset. It turns out the $\|{Q}_i {\boldsymbol{r}}_i\|$ associated with Mg is negligible, due to very localized $2p^6$ electrons of Mg. Nb contributes slightly more than Mg due to more delocalized $4p^6$. In contrast, Wannier centroids associated with Pb and O contribute significantly to $\boldsymbol{P}_{\mathrm{ele}}$ by large off-centering and wide distribution. 
Therefore, we can make the approximation that ${\boldsymbol{r}}_i=0$ for Mg and Nb atoms. This elimination of degrees of freedom reduces the complexity of the model representing ${\boldsymbol{r}}_i$. 
Consequently, we train a Deep Wannier model for Wannier centroids associated with Pb and O with DeePMD-Kit~\cite{zeng2023deepmd} based on the SCAN-DFT data. The data of the first 650 configurations are the training set, and data of the remaining 50 configurations are the testing set. The short-range cutoff of the model is $6$ \AA. The error distributions on the training set and testing set are given by Supplementary Fig.~\ref{Fig:wannier_dist}~(Right). We report an RMSE of 0.12 $\mu \mathrm{C}/\mathrm{cm}^2$ and 0.15 $\mu \mathrm{C}/\mathrm{cm}^2$ on $\boldsymbol{P}_{\mathrm{ele}}$ for the training set and testing set, respectively.

In addition, we introduce an \textit{ad hoc} definition of the local dipole moment associated with the primitive cell. Each primitive cell of an ABO$_3$ perovskite such as PMN can be identified by its central B atom, which is either Mg or Nb in PMN. Let $i$ be the central atom. It is convenient to refer the coordinates of the atoms $j$ in the cell to their coordinates in a nonpolar centrosymmetric reference structure by $\Delta \boldsymbol{R}_{j} \equiv \boldsymbol{R}_j - \boldsymbol{R}_j^{(\text{ref})}$. Similarly, for the WCs, we define $\Delta \boldsymbol{r}_{j} \equiv {\boldsymbol{r}}_{j} - {\boldsymbol{r}}_{j}^{(\text{ref})} + \Delta \boldsymbol{R}_j$. In terms of these relative displacements, the local dipole moment of the $i$-th primitive cell is given by 
$\boldsymbol{s}_i = \sum_{j \in \text{cell-}i} \alpha_j \left( Z_{j} \Delta \boldsymbol{R}_{j} + {Q}_j \Delta {\boldsymbol{r}}_{j}\right)$,
where $\alpha_j$ is a weight factor that accounts for atoms shared between neighboring cells. In our case, $\alpha_j=\frac{1}{8}$ for Pb atoms, $\alpha_j=1$ for Mg or Nb atoms, and $\alpha_j=\frac{1}{2}$ for O atoms.

\section{\label{sec:disorder_conf} Radial correlation function}

In this study, the compositional structure of the PMN simulation cell is modeled as a random mixture of an equal number of chemically ordered regions (CORs) and chemically disordered regions (CDRs), with each region containing $3\sqrt{2}\times3\sqrt{2}\times4$ primitive cells.
Using this protocol, we generate four statistically independent compositional structures.

\begin{figure*}[h!]
    \centering
    \includegraphics[width=0.7\linewidth]{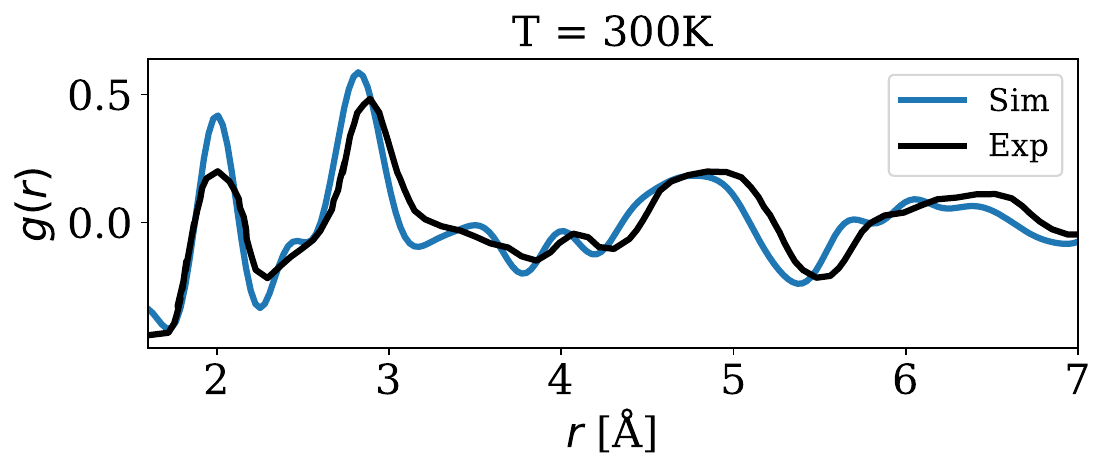}
    \caption{\label{fig:neutron_scattering}The radial pair-correlation function obtained from the calculated total neutron scattering intensity, compared with experiment~\cite{eremenko2019local}.}
\end{figure*}

For each compositional realization, we conducted molecular dynamics simulations in the NVT ensemble at room temperature. The corresponding total radial pair-correlation functions were calculated and subsequently averaged across the four compositional realizations. The resulting profiles are shown in Supplementary Fig.~\ref{fig:neutron_scattering}. The experimentally obtained total radial pair-correlation function is provided for comparison.

The good qualitative agreement between the simulations and the experimental observations~\cite{eremenko2019local} indicates that the adopted composition–structure model adequately represents the key characteristics of compositional disorder in experimental PMN samples.

\section{Distributions and correlations of local dipoles}

We perform NVT molecular dynamics simulations using periodic supercells containing $12\sqrt{2}\times 12\sqrt{2}\times 16$ primitive cells. External electric field is not applied. The system was cooled stepwise from 800 K to 50 K, following the sequence 800, 600, 400, 250, 200, 150, 100, and 50 K, with the 800 K state initialized from an equilibrated paraelectric configuration. Each temperature reduction was implemented with a constant cooling rate over a 100 ps interval.

The simulations were repeated for four independently sampled compositional structure. The resulting trajectories provide an ensemble of atomic configurations, from which local dipole arrangements were subsequently evaluated using the DW model.

\begin{figure*}[tbp]
    \centering
    \includegraphics[width=\linewidth]{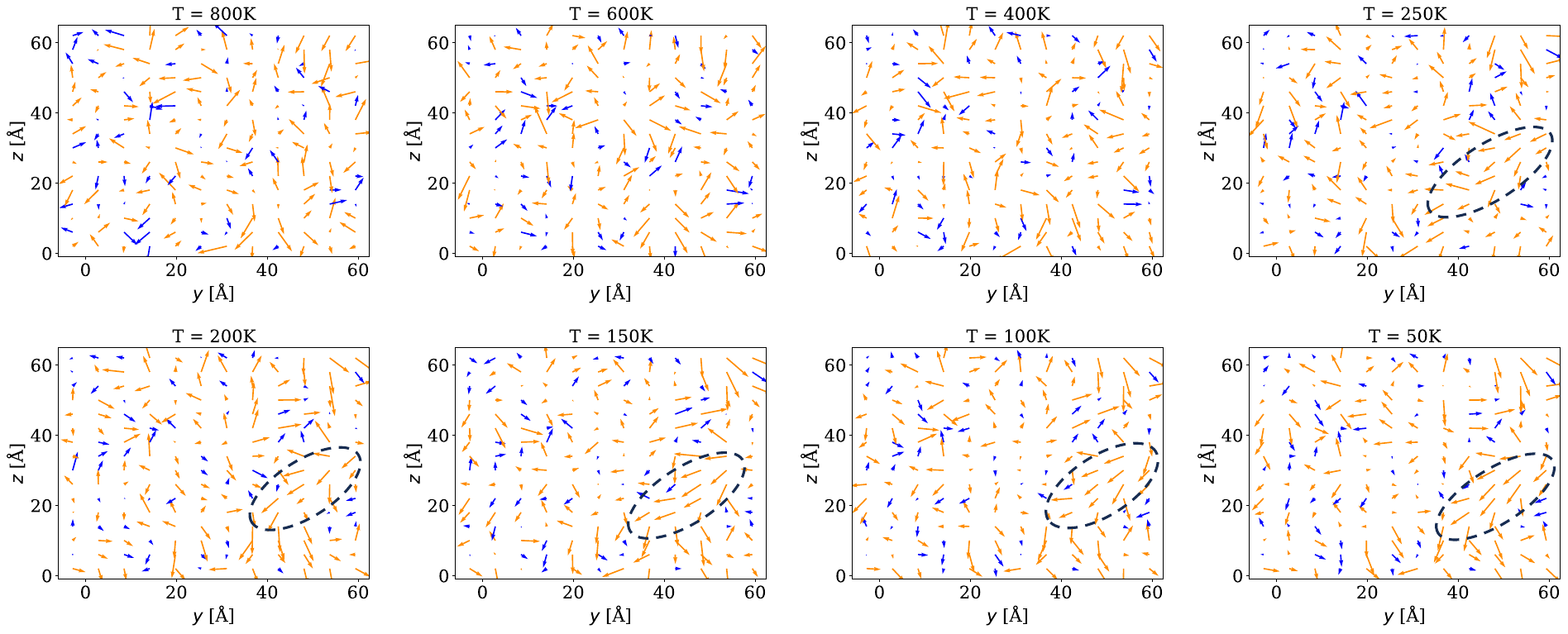}
    \caption{\label{fig:local_dipole_map}Maps of the local dipoles during gradual cooling from 800 K to 50 K.}
\end{figure*}

For a representative compositional structure, Supplementary Fig.~\ref{fig:local_dipole_map} depicts the temperature dependence of the local dipole arrangements, with dipoles within the same unit-cell layer projected onto the [110] plane. Nb-centered dipoles are denoted by orange arrows, whereas Mg-centered dipoles are denoted by blue arrows.

At high temperature, the local dipoles are randomly oriented and continuously fluctuate, but at approximately 250 K, soft PNR with dynamic dipole alignment begin to form in Nb rich environments. Upon further lowering of the temperature, the PNR orientations appear frozen on the nanosecond time scale of the simulations. Nearly identical PNR are obtained by repeating the cooling procedure for the same compositional-disorder realization but starting from a different equilibrium structure at 800 K, indicating that the PNR patterns are dictated primarily by the compositional structure.    

In addition, we calculate the distributions and nearest-neighbor correlations of the local dipole moments, shown in Supplementary Fig.~\ref{fig:dipole_distribution} and~\ref{fig:dipole_correlation}, respectively.  
\begin{figure*}[t]
    \centering
    \includegraphics[width=0.6\linewidth]{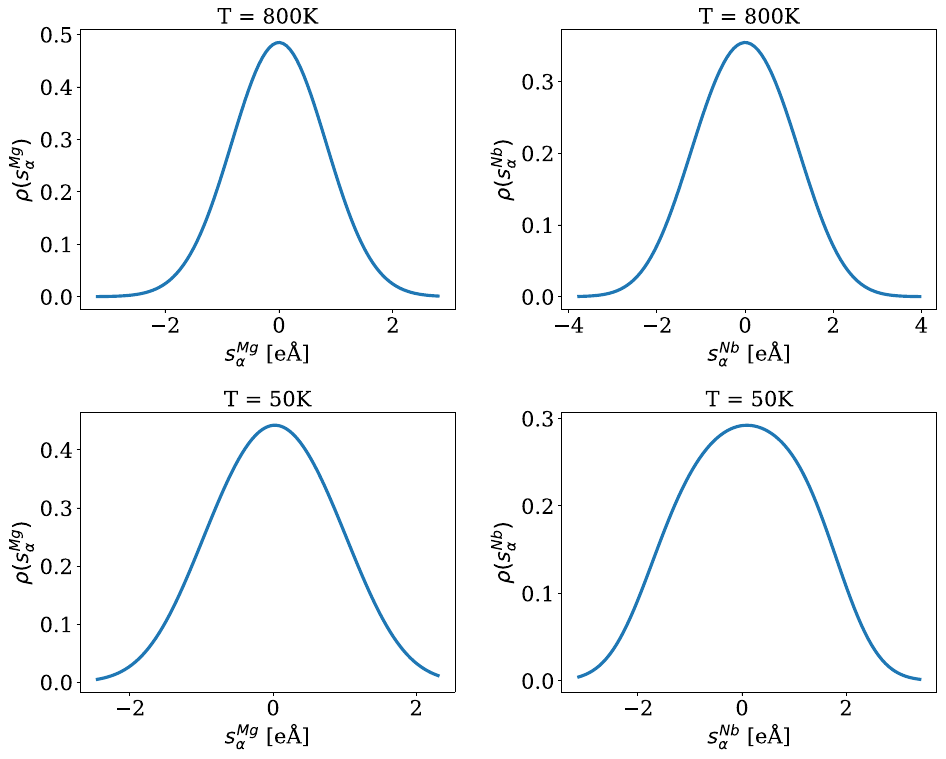}
    \caption{\label{fig:dipole_distribution} Distribution of the $\text{Mg}$- and $\text{Nb}$-dipoles at 800 K and 50 K.}
\end{figure*}
The normalized distributions $\rho(s_{\alpha}^{\mathrm{Nb}})$ and $\rho(s_{\alpha}^{\mathrm{Mg}})$ of the projections of Nb- and Mg-centered local dipole moments along an arbitrary direction $\alpha$ are approximately Gaussian when averaged over the sample, as expected for a system with vanishing macroscopic polarization in the thermodynamic limit. As the temperature decreases, both distributions broaden, reflecting an increase in the magnitude of the local dipoles. Moreover, the variance of $\rho(s_{\alpha}^{\mathrm{Nb}})$ is consistently larger than that of $\rho(s_{\alpha}^{\mathrm{Mg}})$, showing that Nb-centered cells typically carry larger local dipoles.
Supplementary Fig.~\ref{fig:dipole_correlation} further shows that, upon cooling, the distribution of the scalar product between neighboring Nb-centered cell dipoles, $\rho \left( \boldsymbol{s}^{\mathrm{Nb}} \cdot \boldsymbol{s}'^{\mathrm{Nb}} \right)$, becomes increasingly biased toward positive values. This behavior reveals stronger parallel correlations for Nb--Nb dipole pairs than for Mg--Mg or Mg--Nb pairs. Taken together, the larger Nb-centered dipole amplitudes and stronger Nb--Nb nearest-neighbor correlations are consistent with the preferential formation of PNR in Nb-rich regions.

\begin{figure*}[t]
    \centering
    \includegraphics[width=0.7\linewidth]{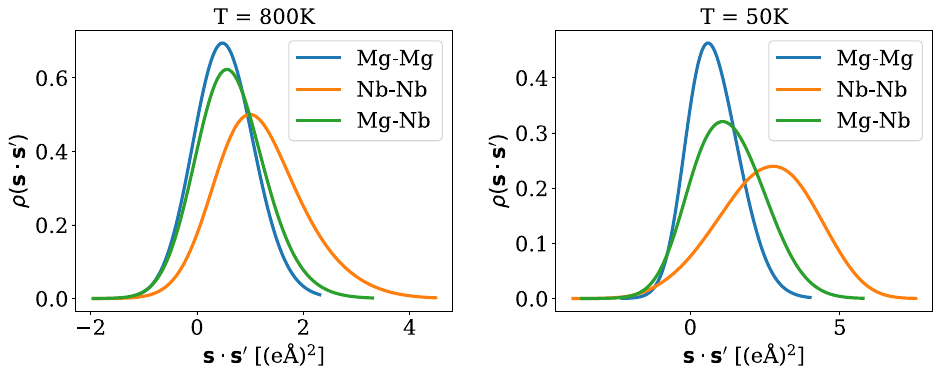}
    \caption{\label{fig:dipole_correlation} Nearest-neighbor correlations of the local dipoles at 800 K and 50 K.}
\end{figure*}

\section{\label{sec:EA_op}Edwards-Anderson order parameter}

The Edwards--Anderson (EA) order parameter is defined as
\begin{equation}
    q = \left[ \frac{1}{N} \sum_i \langle \boldsymbol{s}_i \rangle^2 \right]_{\mathrm{dis}} ,
\end{equation}
where $\langle \cdot \rangle$ denotes a thermal average and $[\cdot]_{\text{dis}}$ stands for an average over compositional-disorder realizations. Originally introduced to characterize the glassy ordering in spin systems~\cite{edwards1975theory, binder1986spin}, the EA order parameter provides a measure of the persistence of local dipoles in the present context.
The temperature dependence of $q$ obtained from our simulations is shown in Supplementary Fig.~\ref{fig:EA_WC_op}(a), where the shaded region indicates the spread among different compositional-disorder realizations.

\begin{figure*}[htbp]
    \centering
    \includegraphics[width=0.35\linewidth]{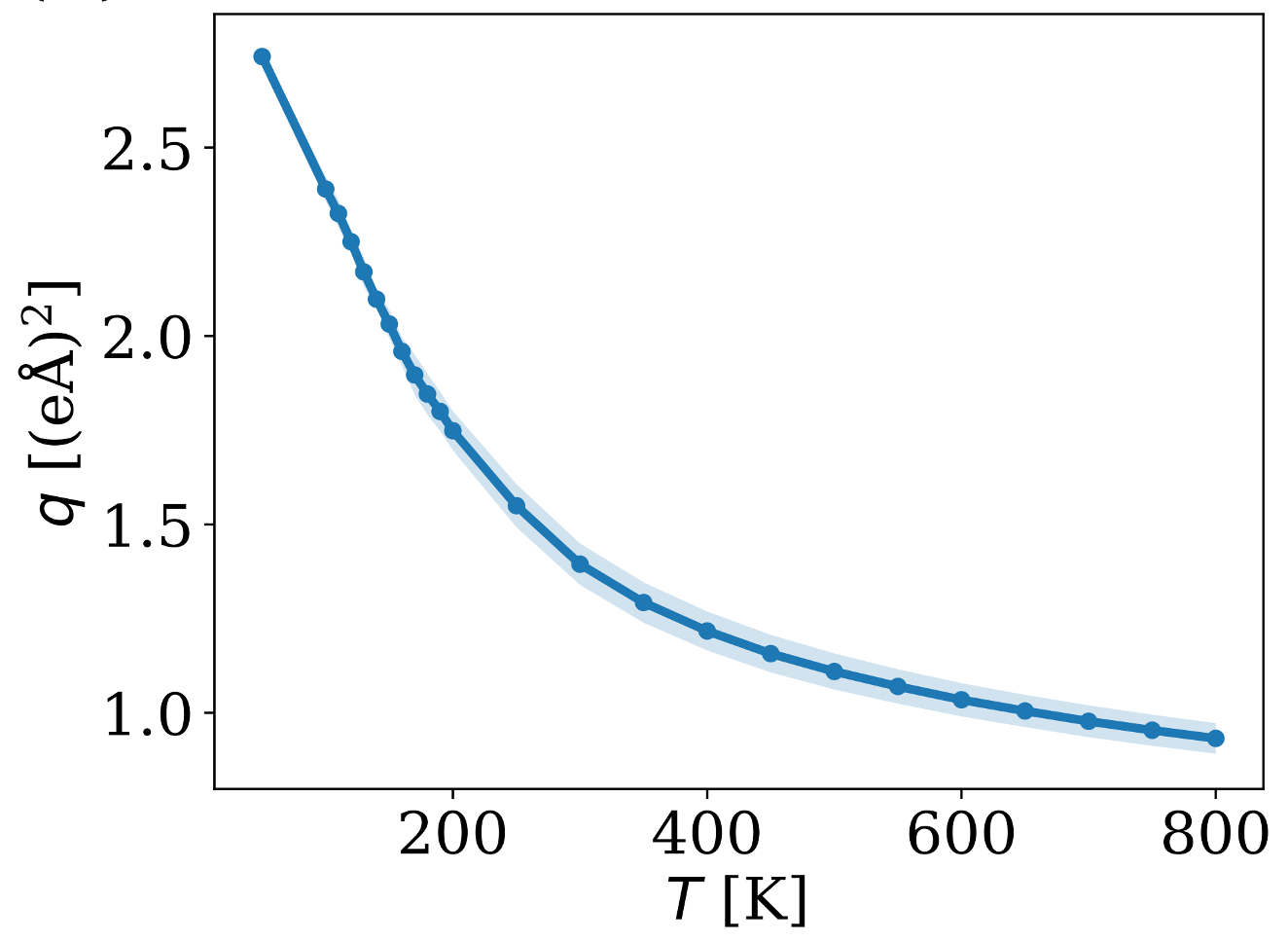}
    \caption{\label{fig:EA_WC_op} Edwards-Anderson order parameter of PMN calculated during the cooling simulation. }
\end{figure*}

The EA order parameter increases gradually as the temperature decreases, indicating the progressive development of glassy order. Interestingly, $q$ remains finite even at elevated temperatures, where the system is in thermal equilibrium, owing to the pinning of local dipoles by compositional disorder. This behavior is analogous to that of spherical spin-glass models with random fields, in which the random fields induce finite local polarization even at high temperature~\cite{pirc1999spherical}. The temperature dependence of $q$ in PMN contrasts with that of conventional ferroelectrics, for which $q$ vanishes in the paraelectric phase, changes sharply across the ferroelectric transition, and saturates at low temperature as the local dipoles align with the spontaneous polarization. A similar gradual evolution of $q$ has also been reported for the relaxor \bzt in Ref.~\cite{akbarzadeh2012finite}.

\section{Vibrational-mode analysis}

\begin{table}[t]
\centering
\caption{\label{tab:PTO_optical_modes}Calculated frequencies (in $\mathrm{cm^{-1}}$) of the zone-center polar optical modes of PTO and the corresponding experimental values from Ref.~\cite{foster1993anharmonicity}. $E$ and $A_1$ denote irreducible representations.}
\begin{ruledtabular}
\begin{tabular}{cccc}
 \multicolumn{2}{c}{This work} & \multicolumn{2}{c}{Ref.~\cite{foster1993anharmonicity}}\\
 $E$&$A_1$&$E$&$A_1$\\ \hline
 77.9 & 160.3 & 87.5 & 148.5 \\
 204.0 & 390.0 & 218.5 & 359.5 \\
 512.8 & 663.1 & 505.0 & 647.0 \\
 287.3 & & 289.0 \\
\end{tabular}
\end{ruledtabular}
\end{table}
The main text presents vibrational-mode analyses of PMN and PTO, including a comparative evaluation of their vibrational densities of states (VDOS) and oscillator strengths.
The Born effective charge tensors used in the calculation of the oscillator strengths are described below. 

\begin{figure*}[b]
    \centering
    \includegraphics[width=0.35\linewidth]{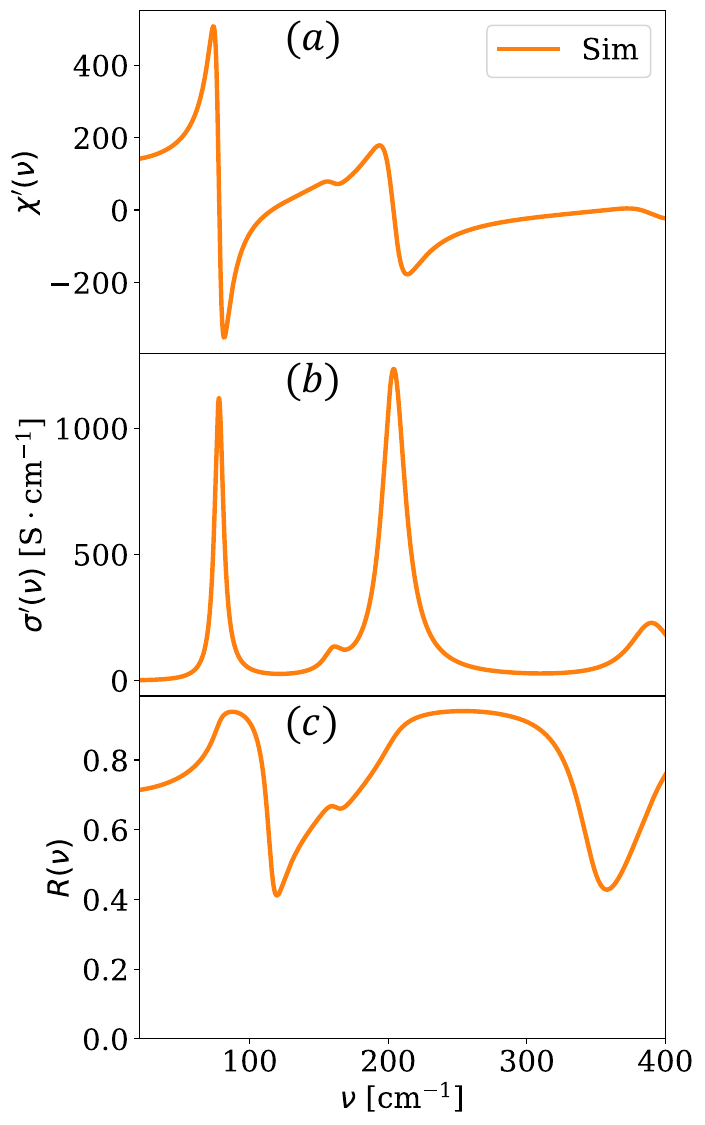}
    \caption{\label{fig:PTO_ha_spectrum} Spectra of (a) dielectric permittivity, (b) optical conductivity and (c) reflectivity of the same PTO supercell.}
\end{figure*}

For PMN, the Born effective charge tensors were based on density functional perturbation theory (DFPT) calculations for a chemically ordered, high-symmetry reference structure of the $P3m1$ space group~\cite{choudhury2006first}. Because atoms of the same chemical species can occupy inequivalent local environments, multiple Born effective charge tensors may occur for a given species. For simplicity, we assume a single species-dependent tensor and therefore average the tensors reported in Ref.~\cite{choudhury2006first} over all crystallographically inequivalent sites of the same chemical species. The resulting tensors are given as
\begin{equation}
\begin{aligned}
    \boldsymbol{Z}^{\mathrm{Pb}} &= \left(\begin{matrix}
        3.66 & -0.0233 & -0.0233 \\
        -0.0233 & 3.66 & -0.0233 \\
        -0.0233 & -0.0233 & 3.66
    \end{matrix}\right) , \quad 
    \boldsymbol{Z}^{\mathrm{Mg}} = \left(\begin{matrix}
        2.85 & -0.01 & -0.01 \\
        -0.01 & 2.85 & -0.01 \\
        -0.01 & -0.01 & 2.85
    \end{matrix}\right) \\
    \boldsymbol{Z}^{\mathrm{Nb}} &= \left(\begin{matrix}
        6.625 & -0.23 & -0.23 \\
        -0.23 & 6.625 & -0.23 \\
        -0.23 & -0.23 & 6.625
    \end{matrix}\right) , \quad 
    \boldsymbol{Z}^{\mathrm{O}} = \left(\begin{matrix}
        -2.7 & 0.103 & -0.0967 \\
        0.0533 & -2.9 & 0.06 \\
        0.02 & 0.227 & -3.427
    \end{matrix}\right) .
\end{aligned}
\end{equation}

For PTO, the Born effective charge tensors were based on DFPT calculations for the relaxed tetragonal cell of $P4mm$ space group~\cite{kuma2019structural}. In particular, the oxygen atoms occupy three inequivalent local environments and are therefore associated with three distinct tensors. Consistent with the same species-dependent approximation adopted for PMN, we average these oxygen tensors and assign a single effective tensor to each chemical species for the tetragonal cell. The resulting tensors are
\begin{equation}
    \boldsymbol{Z}^{\mathrm{Pb}} = \left(\begin{matrix}
        3.85 & 0 & 0 \\
        0 & 3.85 & 0 \\
        0 & 0 & 3.63
    \end{matrix}\right) , \quad 
    \boldsymbol{Z}^{\mathrm{Ti}} = \left(\begin{matrix}
        7.77 & 0 & 0 \\
        0 & 7.77 & 0 \\
        0 & 0 & 6.55
    \end{matrix}\right) , \quad
    \boldsymbol{Z}^{\mathrm{O}} = \left(\begin{matrix}
        -3.74 & 0 & 0 \\
        0 & -3.74 & 0 \\
        0 & 0 & -3.27
    \end{matrix}\right) .
\end{equation}

The approximations applied to the Born effective charge tensors do not modify the qualitative characteristics of the oscillator-strength spectrum, in particular the pronounced contrast observed between PTO and PMN.

The main text has shown that PTO exhibits a few zone-center optical modes that are infrared (IR) active. Here, we present the complete set of zone-center polar optical modes in PTO, as obtained from our first-principles model. These modes are summarized in Table~\ref{tab:PTO_optical_modes}, along with the polar optical modes determined experimentally from spectroscopic measurements~\cite{foster1993anharmonicity}. The comparison indicates quantitative agreement between the model predictions and the experimental results.

Then, we adopt the damped harmonic-oscillator approximation and calculate the dielectric susceptibility $\chi(\nu)$ of PTO using Eq. (3) in the main text. The real part of the optical conductivity ($\sigma'$) and the optical reflectivity ($R$) of PTO can be consistently determined. The results are reported in Supplementary Fig.~\ref{fig:PTO_ha_spectrum}. 
In contrast to PMN, PTO shows vanishing phononic contribution to the optical conductivity in the low-frequency limit. This is expected because PTO has no IR-active vibrational modes at sufficiently low frequencies, while PMN has excess low-frequency polar vibrational modes.

In PMN, vibrational modes with frequencies below approximately 20 $\mathrm{cm}^{-1}$ are localized on relatively small groups of atoms. These quasi-localized excitations (QLE) are distributed throughout the simulation cell, as shown by the representative example in Supplementary Fig.~\ref{fig:ha_qle_cores_gr}(a). The  distribution of inter-core distances between QLE is reported in Supplementary Fig.~\ref{fig:ha_qle_cores_gr}(b). The corresponding radial distribution function, $g(r)$, is shown as a histogram in Supplementary Fig.~\ref{fig:ha_qle_cores_gr}(c). Its resolution is limited by the restricted available statistics.
\begin{figure*}[hbtp]
    \centering
    \includegraphics[width=\linewidth]{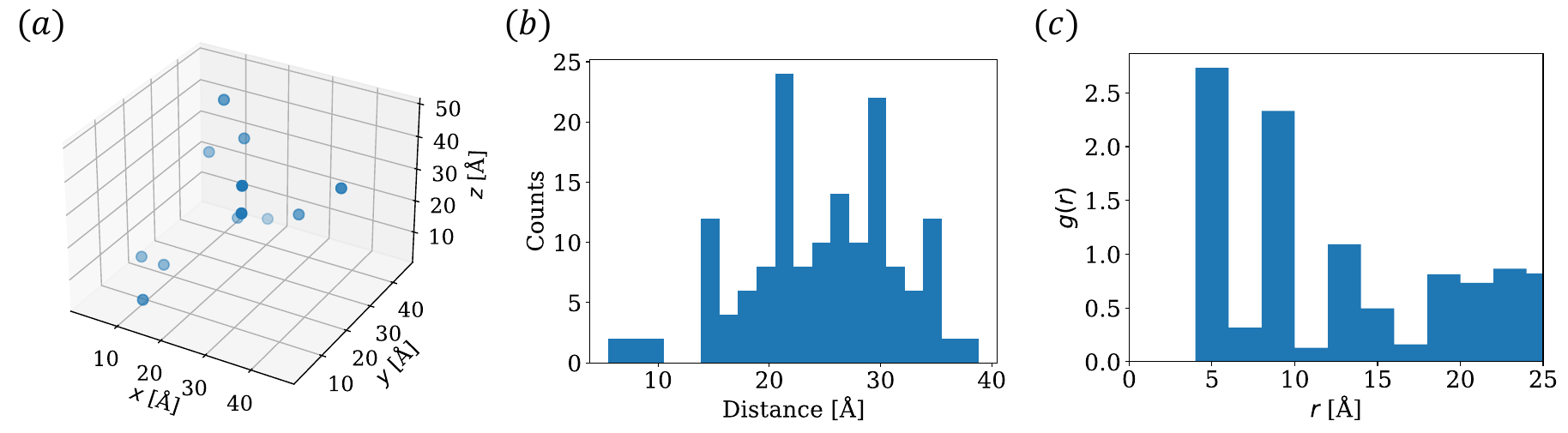}
    \caption{\label{fig:ha_qle_cores_gr} (a) The visualization of centered cores of several QLE modes. (b) The distribution of inter-core distances between QLE. (c) The radial distribution function $g(r)$ of QLE cores.}
\end{figure*}

\clearpage
\bibliography{ref}